# Systematic Modulation of Charge and Spin in Graphene Nanoribbons on MgO

Amelia Domínguez-Celorrio[1,2,3,†], Leonard Edens[4, †], Sofía Sanz[5], Manuel Vilas-Varela[6], Jose Martinez-Castro[7], Diego Peña[6], Véronique Langlais[8], Thomas Frederiksen[5,9], José I. Pascual[4,9], and David Serrate[1,10,11*]

[1]Insituto de Nanociencia y Materiales de Aragón (INMA), CSIC-Universidad de Zaragoza, Zaragoza, E-50009, Spain
[2]School of Physics and Astronomy, Monash University, Clayton, VIC 3800, Australia
[3]ARC Centre for Future Low Energy Electronics Technologies, Monash University, Clayton, VIC 3800, Australia
[4]CIC NanoGUNE BRTA, San Sebastián, E-20018, Spain
[5]Donostia International Physics Center, San Sebastián, E-20018, Spain.
[6]Centro Singular de Investigación en Química Bilóxica e Materiais Moleculares (CiQUS) and Departamento de Química Orgánica, Universidade de Santiago de Compostela, Santiago de Compostela, E-15782, Spain.
[7]Peter Grünberg Institut (PGI-3), Forschungszentrum Jülich, 52425 Jülich, Germany
[8]Centre d'Elaboration de Materiaux et d'Etudes Structurales, CNRS, Toulouse, F-31055 France
[9]Ikerbasque, Basque Foundation for Science, Bilbao, E-48013, Spain.
[10]Departamento de Física de la Materia Condensada, Universidad de Zaragoza, Zaragoza, E-50009, Spain
[11]Laboratorio de Microscopias Avanzadas (LMA), Universidad de Zaragoza, Zaragoza, E-50018, Spain.

[†]These authors contributed equally
*email: serrate@unizar.es

**Graphene nanostructures can be engineered with atomic precision to display customized electronic states with application in spintronics or quantum technologies. In order to take advantage of their full potential, their charge and spin state must be precisely controlled. Graphene systems exchange charge to reach thermodynamic equilibrium with their environment, requiring external gating potentials to tune their ground state. Alternative strategies like intrinsic doping or substrate modifications provided small variations of their equilibrium charge and poor control over their spin. Here, we show systematic manipulation of the electron occupation in graphene nanoribbons (GNRs) laying on MgO layers grown on Ag(001). Owing to the extraordinary decoupling properties of MgO, and the electropositive character of the substrate, GNRs are found to host an integer number of electron charges that depend on their length and shape. This results in the alternation between a non-magnetic closed-shell state and an open-shell paramagnetic system for even and odd electron occupations respectively. For the odd case, we found the spectral fingerprint of a narrow Coulomb correlation gap, which is the smoking gun of its spin ½ paramagnetic state. Comparisons of scanning tunnelling microscopy (STM) data with mean-field Hubbard (MFH) simulations confirm the practical discretization of the GNR electronic states and point to charge excess of up to 19 electrons in a single ribbon. We anticipate that GNRs supported by MgO ultra-thin insulating films can open the door to customized devices for quantum sensing and quantum processing applications.**

## INTRODUCTION

Establishing parameters to systematically manipulate the charge state of polycyclic aromatic hydrocarbons is crucial to harness their electrical, chemical and magnetic functionalities. On-surface synthesis (OSS) techniques [1–3] allow us to manufacture fixed configurations of the conjugated π-electron cloud with designated purposes in nanographenes, but with no control over their post-synthesis electronic occupancy. A paramount example is the effort to trigger the appearance of open-shell magnetic states as a response to the enhanced electronic correlations [4]. Such states are often spatially localized and can be realised by enforcing particular shapes or edge geometries [5–14], or by defects in the $sp^2$ lattice [15,16]. Nanographenes generally lie on a metallic surface after synthesis, which leads to the screening of electron correlations due hybridization effects. On metallic surfaces, small structures can nevertheless display robust intrinsic magnetism due to correlations splitting [5–10,13]. On the contrary, extended structures like graphene nanoribbons (GNR) have shown either no magnetic fingerprints [12,17,18] or, at most, extremely small correlation gaps of about ~10 meV accompanied by a weak spin polarization [19].

By electrically decoupling the GNRs or other nanographenes from the catalysing metal, we can restore electronic correlations and preserve the quantum behaviour associated to their electronic and magnetic degrees of freedom. In addition, the use of insulating spacers is also expected to suppress relaxation channels of intrinsic electronic and spin quantum states, maintaining the long electron-spin coherence times of free standing carbon-based quantum dots [20,21]. To this end, previous studies explored the intercalation of ultrathin insulators after growing arm-chair GNRs [22,23]. Although the discretization gaps could be measured with much better energy resolution, the electronic decoupling is not enough as to induce any correlations splitting. The end states of arm-chair GNRs have been also assessed on NaCl bilayers on Au(111) [24] and $TiO_2$-rutile [25], in this case finding correlation gaps at the zig-zag termini very close to the theoretical predictions. Finally, GNRs with longitudinal edges dense in zig-zag segments have been lifted from the Au(111) reactive surface and dropped onto NaCl bilayers [12,26], finding robust correlation gaps just for the case of pure zig-zag edges [12]. None of these decoupling strategies have provided a way to manipulate the number of singly occupied molecular states. Moreover, the molecular resonances reported so far on decoupling layers exhibit broad linewidths of at least 100 meV, suggesting that these systems are not free from hybridization channels, which is detrimental to the behaviour as spin quantum dots. In fact, it is known that NaCl introduces coupling with phonons [27] and electronic bands [28,29].

Inspired by the excellent results of ultra-thin MgO layers as passive support for magnetic atoms and molecules [30–32], here we employ MgO monolayers ($MgO_{ML}$) grown on Ag(001). Charge transfer from Ag(001) across the MgO takes place in integer number of electrons, which permits to control the odd/even electron occupation of the QW states by means of the GNR length. The addition of a single molecular precursor unit (PU) can affect the odd/even occupancy of the GNR. The MgO spacer also enhances electron-electron (e-e) interactions that stabilize odd occupations with correlation gaps of a few tens of meV. This is the signature of a spin-1/2 state associated to odd numbers of electrons in the mean-field-Hubbard (MFH) approximation. The quantum-well (QW) states confined at the longitudinal edges of finite GNRs (2 to 12 nm) on $MgO_{ML}$ display an exceptionally small intrinsic linewidth of 1 meV.

## RESULTS AND DISCUSSION

Chiral graphene nanoribbons (*m,n,w*)-GNRs grow with their longitudinal axis along a vector $(m,n)$ of the graphene lattice, enclosing *w* C-C pairs among them. We work with (3,1,8)- and (3,2,8)-GNRs, whose edge consists in an alternating sequence of three zig-zag graphene vectors and arm-chair segments (Fig. 1a). As a consequence, they exhibit edge states and a symmetry-protected topological (SPT) energy gap, giving rise to zero-energy SPT end states at their termini [17]. In the presence of electronic correlations, both edge and SPT states can host spins with different degree of delocalization. However, on the rather electrophilic substrate like Au(111) and the related intermetallic surface $GdAu_2$, we found slight doping of holes and electrons respectively [17,19], with negligible correlation gaps.

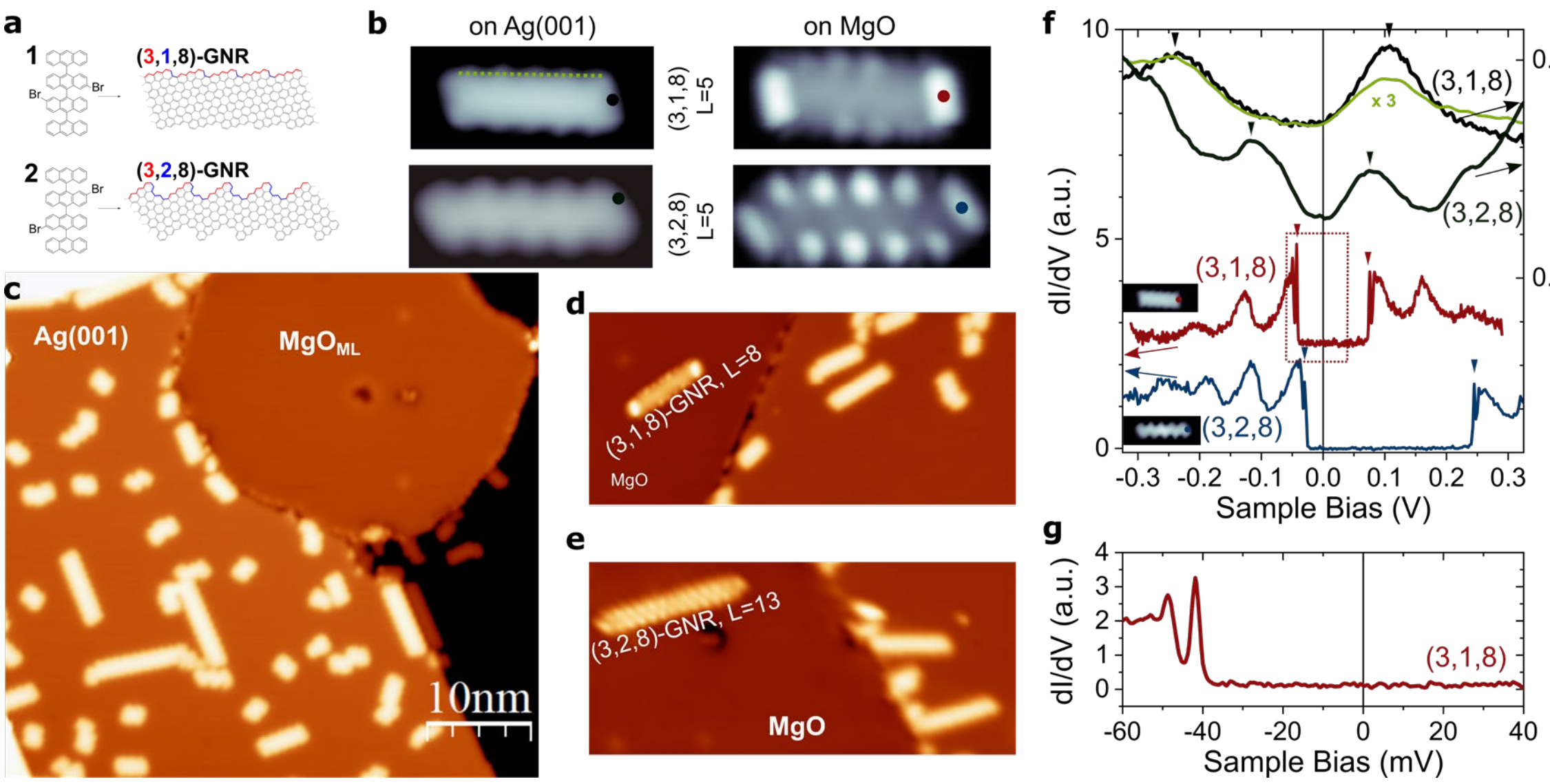

**Figure 1.- STM characterization of (3,*n*,8)-GNRs on MgO/Ag(001). a)** Chemical structures of the precursors and the resulting (3,1,8)- and (3,2,8)-GNRs. **b)** STM topography ($V_b$ =0.5 V, $I_t$ = 10 − 70 pA, image width 6 nm) of (3,1,8) (upper row) and (3,2,8) (bottom row) GNRs with *L*=5 precursor units length on Ag(001) and on $MgO_{ML}$/Ag(001). **c)** STM image ($V_b$=0.5 V, $I_t$=50 pA) of a $MgO_{ML}$ island coexisting with GNRs synthesized over the regions of bare Ag(001). **d,e)** Instances of successful atomic manipulation of GNRs from the Ag surface to the MgO island ($V_b$=0.5 V, $I_t$=50 pA). **f)** *dI/dV* spectra of the *L*=5 (3,1,8) and (3,2,8)-GNRs on Ag(001) and on the $MgO_{ML}$ (stabilization $V_b$=0.5 V, $I_t$=100 pA, $V_{mod}$=1 mV and 8 mV r.m.s. for spectra on MgO and Ag(001) respectively). The colour code indicates the positions in panel (b) where the spectra were acquired. Insets show the in-gap ($V_b$ =5 mV) constant-height tunnelling current image of the GNRs on MgO. The green curve is an average of spectra taken along the dashed line in panel (b). **g)** High resolution *dI/dV* spectrum within the region enclosed by the dotted rectangle in (f) (stabilization $V_b$ =0.5 V, $I_t$ =200 pA and $V_{mod}$ =0.5 mV r.m.s.).

We synthesized (3,*n*,8)-GNRs on Ag(001) by thermally activated Ullmann coupling and subsequent cyclodehydrogenation of precursors **1** and **2** [17,19], leading to GNRs with *n*=1 and 2, respectively (see Fig. 1a,b and Supplementary Fig. S2) with varying lengths between *L*=2 to 22 precursor units (PU). High-resolution constant height current images (Supplementary Fig. S2) with CO-functionalized tips confirm the correct structure of the GNRs. Differential conductance (*dI/dV*) spectra on the GNRs over the Ag(001) surface display a set of broad peaks around the Fermi level (Figure 1f) that, in anticipation to our later results, can be assigned to QW states stemming from the confinement of the conduction band localized at the edges [17,19]. The typical full-width at half-maximum (FWHM) of these resonances varies from ~50 meV to 100 meV (see Fig. 1f and Supplementary Fig. S4).

The GNRs were transferred to the previously grown MgO monolayer patches by means of lateral manipulation with the STM tip (see Fig. 1c-e and Methods). This technique involves smaller forces and is less invasive than vertical manipulation mode that is necessary to relocate GNRs onto other insulators like NaCl. Lateral manipulation is possible because the MgO ML islands are embedded on the Ag surface, and thus GNRs are synthesized right on the plane of the MgO surface (see Supplementary Fig. S3). The GNRs over the MgO island exhibit a distinct change in STM topographic images (Fig. 1b) and, more importantly, their electronic structure changes drastically. As shown in Fig. 1f,g for (3,n,8)-GNRs with $L$=5 PU, the molecular resonances are much sharper, featuring linewidths of the order of 1 meV (Supplementary Fig. S5 illustrates how it is determined). Each resonance appears followed by two satellite peaks at 7.6 and 76 mV distance (independently of the GNR length) that correspond to the excitation of Frank-Condon (FC) resonances (Supplementary Fig. S6 and S11), in analogy with previous studies of individual molecules and GNRs weakly coupled to metal surfaces [33–35].

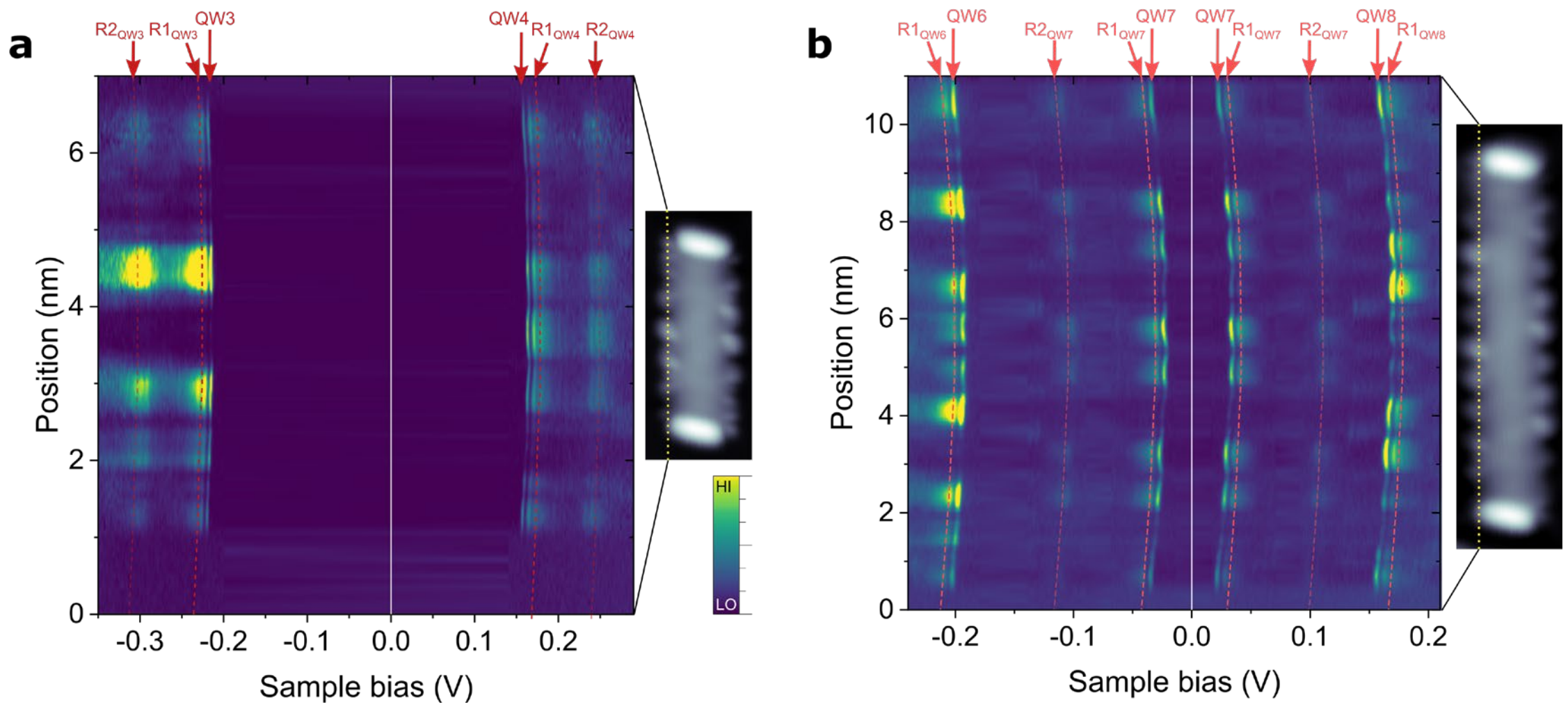

**Figure 2.- Detailed electronic structure of edge state in GNRs/MgO.** Stack plots of *dI/dV* spectroscopy curves taken along the chiral edge of (3,1,8)-GNRs with $L$=6 **(a)** and $L$=11 **(b)**. Insets show topography images ($V_b$=0.5 V, $I_t$=50 pA). Dotted yellow lines indicate the positions where spectroscopy was retrieved. Spectroscopy parameters: stabilization $V_b$=0.5 V and $I_t$=200 pA, $V_{mod}$=1 mV rms, $T$=1.2 K and 4.3 K for panel (a) and (b) respectively. n$^{th}$ order QW states are labelled as QWn. Reddish dashed lines are a guide to the eye of the position dependent energy of their associated Frank-Condon resonances ($Rm_{QWn}$: m$^{th}$ replica of the n$^{th}$ QW state, see Supplementary Fig. S3b).

To elucidate the orbital character of the sharp GNR resonances on MgO, we acquired *dI/dV* spectroscopy curves along the GNRs edges. Figs. 2a and 2b show stacked spectroscopy plots as a function of bias and position along the edge for (3,1,8)-GNRs with $L$=6 and 11, respectively. In $L$=6 (Fig. 2a and 3a) we observe two QW states (each one flanked by their Frank-Condon replica) separated by a large energy gap of 380 meV. Their spatial distribution shows conductance maxima distributed along the edge, with different number and position of nodal planes on either side of the Fermi level, as would be the case for a discretization gap of the conduction band. In the case of $L$=11, there are four different QW states within the same energy window (Fig. 2b and 3b), what is consistent with the larger GNR length. Notably, for $L$=11, the frontier QW states around the Fermi level exhibit identical intensity profiles along the edge. As will be discussed below, this feature is characteristic of a correlation gap, as opposed to the discretization gap proposed for $L$=6.

We performed a quantitative assignment of QW order to the resonances on both Ag(001) and $MgO_{ML}$ by comparing in Fig. 3a,b their experimental spatial distribution with MFH calculations of the local density of states (LDOS) (see Methods). As shown in Extended Data Fig. 1, due to the intrinsic peak broadening on the metal, the QW states close to the SPT state with small energy spacing cannot be individually resolved. Still, the position of the SPT state can be tracked down as a broad occupied peak, which lies well below the Fermi level on Ag(001). On the other hand, the QW states around the Fermi level can be readily identified. For $L$=6 (3,1,8)-GNR on Ag(001), the experiments resolve an occupied state at -42 mV and a fully unoccupied state at 245 mV (Fig. 3a, left). MFH results for the LDOS of the 3rd and 4th QW states reproduce the experimental $dI/dV$ maps of both peaks (Fig. 3a, right). Taking into account that each QW state can accommodate two electrons (with opposite spin), plus two additional electrons in the SPT state (one at each ribbon end), we conclude that this ribbon hosts a fractional charge state of slightly less than $q$=8 $e^-$ excess electrons relative to the charge-neutral case. Note that the 3rd QW state is still partially unoccupied due to Fermi level pinning. When the GNR is transferred onto the MgO island, all states experience an energy downshift, whereas their intensity distribution still corresponds to the LDOS of the same QW states (Fig. 3a). The 3rd QW state shifts down to -216 mV, and the 4th QW state appears at 163 mV. Therefore, the GNR retains approximately the same occupancy, although in this case, the charge doping is exactly $q$=8 $e^-$, because there is no in-gap spectral weight. As sketched in Fig. 3a, the n-doping on the oxide layer is a consequence of the reduced work function of MgO/Ag(001) relative to the bare Ag(001) [36], which we have determined as $\delta\Phi_{[Ag\text{-}MgO]}$=0.63 eV from the analysis of their respective field emission resonances (see Supplementary Fig. S7).

The extremely narrow lineshape of the molecular resonances on MgO, and the appearance of FC resonances that require transitions to long-lived excited molecular states [33], points to a very efficient electronic decoupling from the metal substrate. In this context, one would expect enhanced e-e interactions [24,37]. This is the case of the (3,1,8)-GNR with $L$=11 shown in Fig. 3b, which displays a qualitatively different behaviour than the $L$=6 case. On Ag(001) we find a partially occupied state at -23 mV and a fully unoccupied state at 123 mV. MFH simulations (see also Supplementary Fig. S8) reveal that they correspond to the 6th and 7th QW states. Taking into account the two additional electrons hosted by the SPT state, we conclude a charge state of slightly less than $q$=14 $e^-$. On MgO, all states experience again an energy downshift. The 6th QW state shifts in energy down to -200 mV. The 8th QW state, which was barely resolved on Ag(001) at around 250 mV, is now prominent at 171 mV. Remarkably, the 7th QW appears split in two states with identical spatial distribution at -30 mV and 27 mV. Since only one of them is occupied, we conclude an odd occupancy of exactly $q$=15 $e^-$, which implies one singly occupied and one singly unoccupied frontier state with opposite spin, as depicted by the arrows in Fig. 3b. This remarkable large negative charge is in line with the huge energy shift of the SPT state of -650 meV, which would mark the charge neutrality point in the absence of charge transfer from the substrate (see Extended Data Fig. 2 and the discussion about high bias range spectroscopy in Supplementary Section III).

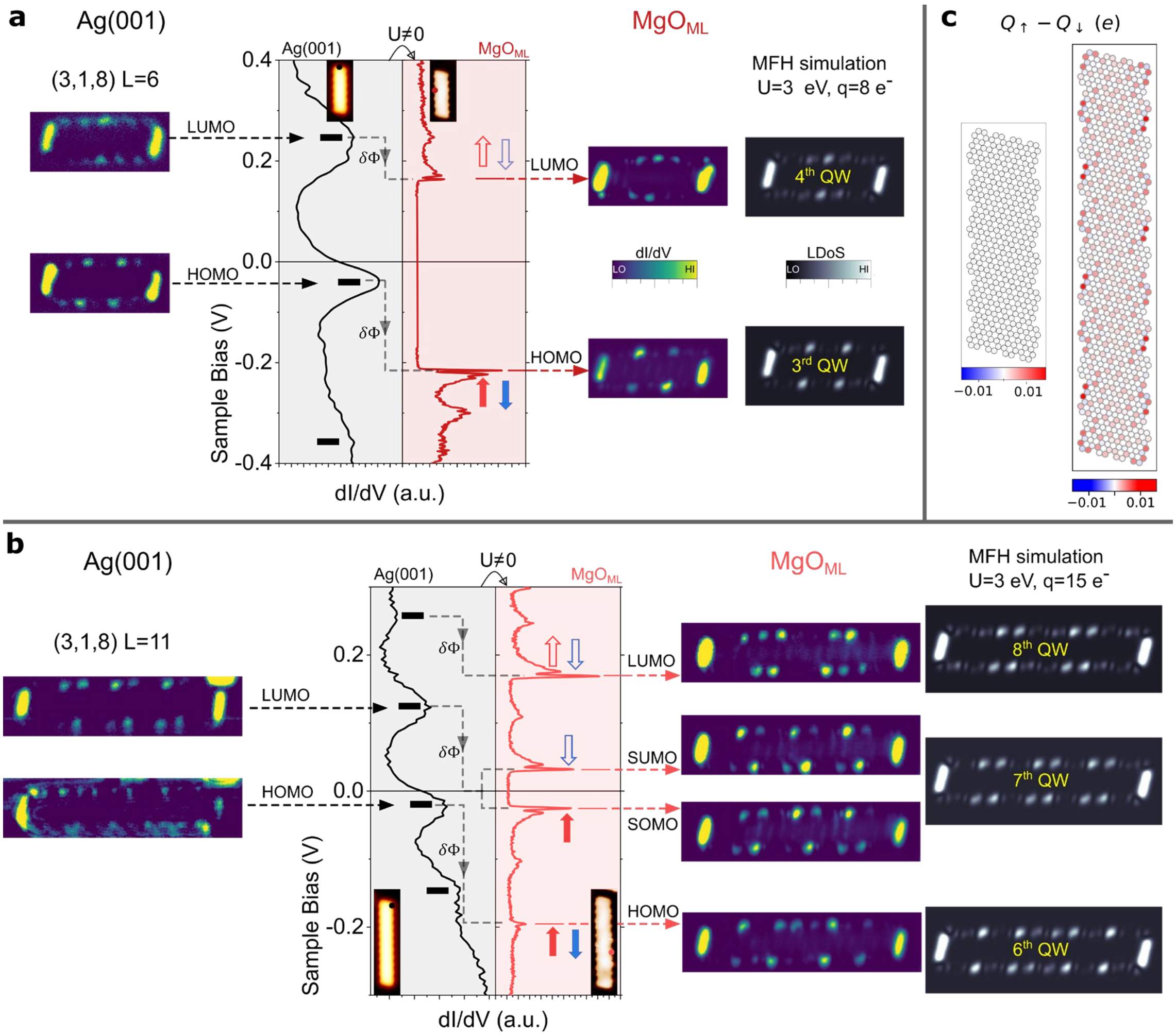

**Figure 3.- Even/odd occupancy of discrete QW states in (3,1,8)-GNRs. (a,b)** *dI*/dV point spectra of L=6 and 11 GNRs respectively taken at the positions marked in the insets. Images on the left column are experimental constant current *dI/dV* maps at the indicated energies on Ag(001). Images on the right column are constant height *dI/dV* maps on $MgO_{ML}$ and theoretical LDOS simulations (see Methods) of the QW states indicated by the yellow labels. Black[red] dashed arrows indicate the energy at which the different QW states appear on Ag(001)[$MgO_{ML}$], connecting the spectroscopic features with their corresponding maps. For *L*=11 (b), the 6$^{th}$ QW state is close enough to the Fermi level so that the finite *U* plays a significant role, splitting the two spin channels. The red/blue filled[empty] arrows represent the spin of the single electron occupied[unoccupied] states. **(c)** Calculated MFH spin polarization projected on the C sites (see Methods) for *L*=6 and *L*=11 GNRs with *U*=3.0 eV and charge states of 8 and 15 electrons, respectively. STM parameters: Insets in (a) and (b) display the topography at $V_b$=0.5 V and $I_t$=50 pA in the case of Ag(001) and constant height in-gap current at $V_b$=2 mV in the case of $MgO_{ML}$. *dI/dV* spectroscopy parameters in (a) and (b): stabilization at 0.5 V/100 pA ($V_{mod}$=5-8 mV) for Ag(001) and 0.5 V/200 pA ($V_{mod}$=1 mV) for $MgO_{ML}$. Parameters for *dI/dV* mapping: on Ag(001) constant current maps are taken at $V_b/I_t$=210 mV/200 pA (QW4) and -72 mV/200 pA (QW3) with $V_{mod}$=5 mV for *L*=6 (a), and 123 mV/100 pA (QW7) and -30 mV/100 pA (QW6) with $V_{mod}$=8 mV for *L*=11 (b); constant height maps on $MgO_{ML}$ are taken after opening the feedback at the GNRs' centre with regulation set points of -200 mV/200 pA (QW4) and 150 mV/150 pA (QW3) with $V_{mod}$=1 mV for *L*=6 (a), and of 50 mV/20[100/5] pA (QW8[QW7/QW6]) with $V_{mod}$=1 mV for *L*=11 (b). *T*=1.2 K

The work function reduction in the MgO ($\delta\Phi_{[Ag-MgO]}$, see Supplementary Fig. S7) acts here as a gating potential for the QW states (Supplementary Table S1 and Fig. S9) [36]. We represent its associated energy downshift in Figs. 3a,b by dashed grey vertical arrows. Many-body correlations are taken into account in the MFH model by the on-site Coulomb repulsion, *U*. Its effect for the 7$^{th}$ QW state of *L*=11 at half-filling as *U* increases is to induce a splitting of the two spin channels (experimental value in Fig. 3b of ~50 meV). Due to the confinement of the flat edge state band [17,19], if the 7th QW state is sufficiently close to the Fermi level for $U$=0, small $U$ values will cause the population of one additional state with one electron from the Ag(001) reservoir. Within the MFH model, this leads to a total spin-½ distributed as depicted by the simulation of the spin density shown in Fig. 3c. In contrast, for $L$=6, electron correlations ($U$>0) renormalize the eigenenergies (~10 meV for $U$=3 eV), but the highest occupied molecular state remains deep below the Fermi level, making the double occupancy (i.e. closed shell configuration) energetically favourable and rendering a non-magnetic ground state.

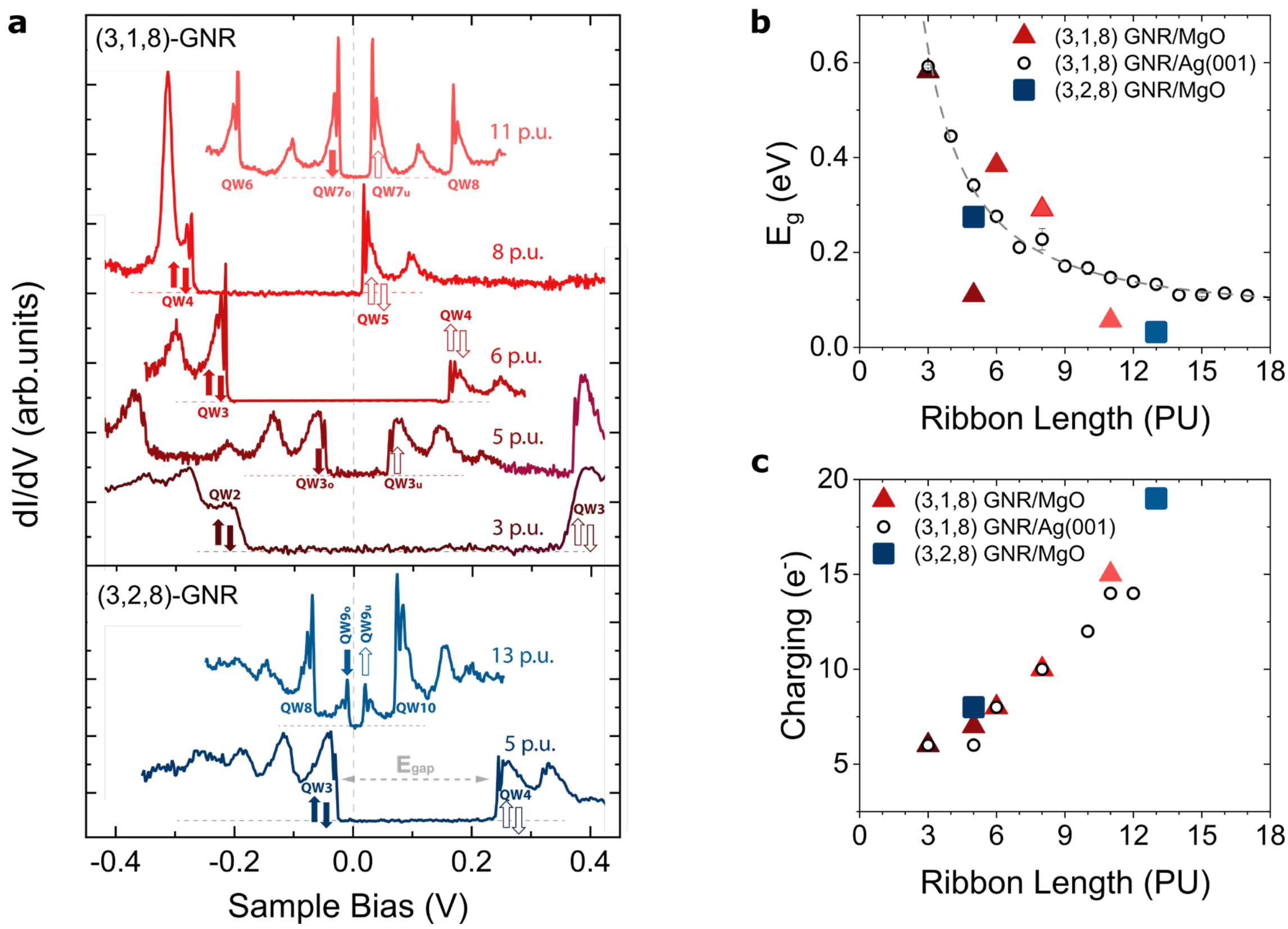

**Figure 4.- Energy gap across the Fermi level in (3,*n*,8)-GNRs repositioned on MgO$_{ML}$. a)** Waterfall plot of $dI/dV$ spectra (stabilization at $V_b$ =0.5 V, $I_t$ =200 pA; $V_{mod}$ =1 mV r.m.s.; $T$=1.2 K for $n$=1 and 4.3 K for $n$=2) of several (3,$n$,8)-GNRs showing the non-monotonous behaviour of the gap across the Fermi level. The acquisition position in the case of $L$=6,11 is shown in the insets of Figs. 3a,b. Filled/empty arrows represent occupied/unoccupied single electron states with well defined $S_z$ quantum number. For L=8, the broad peak at -0.3 V is ascribed to the charging of a point defect in proximity (Supplementary Fig. S11). **b)** Evolution of the gap as a function of the GNR length of (3,1,8)-GNRs on Ag(001) –empty circles, error bars are derived from measurements in 2 or 3 GNRs of the same length– and several representative examples of (3,1,8)/(3,2,8)-GNRs on MgO$_{ML}$ –triangles/squares–. The dashed line is an asymptotic fit proportional to $L^{-3/2}$. **c)** Exact charge state of (3,1,8)/(3,2,8)-GNRs on MgO monolayer and approximate charge state of (3,1,8)-GNRs on Ag(001), see Supplementary Table S1 for further details

All (3,1,8)- and (3,2,8)-GNRs investigated here can be classified into the two categories shown in Figure 3, i.e., either odd occupation with identical QW states around the Fermi level, or even occupation with different QW states (see Supplementary Fig. S8). Figs. 4a and 4b illustrate an interesting anomaly in the gap across the Fermi level ($E_g$) of (3,$n$,8)-GNRs on MgO: its size does not vary monotonically as a function of length. For instance, the gap for $n$=1 and $L$=5 is about 100 meV, whereas for just one more PU ($L$=6) the gap increases abruptly to ~400 meV, to experience again a substantial drop between $L$=8 ($E_g$~300 meV) and $L$=11 ($E_g$ ~50 meV). This is in contrast with the evolution of the GNRs' gap on Ag(001), which follows the expected asymptotic decay with increasing length of a particle-in-a-box (Fig. 4b and Supplementary Fig. S4). On MgO, some (3,$n$,8)-GNRs exhibit similar gaps to the ones on Ag(001), while others have a much reduced value. Fig. 4c shows the excess charge ($q$) in all GNRs determined by means of the same procedure as for $L$=6 and $L$=11 GNRs above. The anomalously small values of $E_g$ (Figs. 4b,c) are univocally linked to frontier states with identical spatial distribution (see Supplementary Fig. S8) and, hence, to odd occupations. These GNRs have, thereby, open-shell character of the edge state and non-zero total spin.

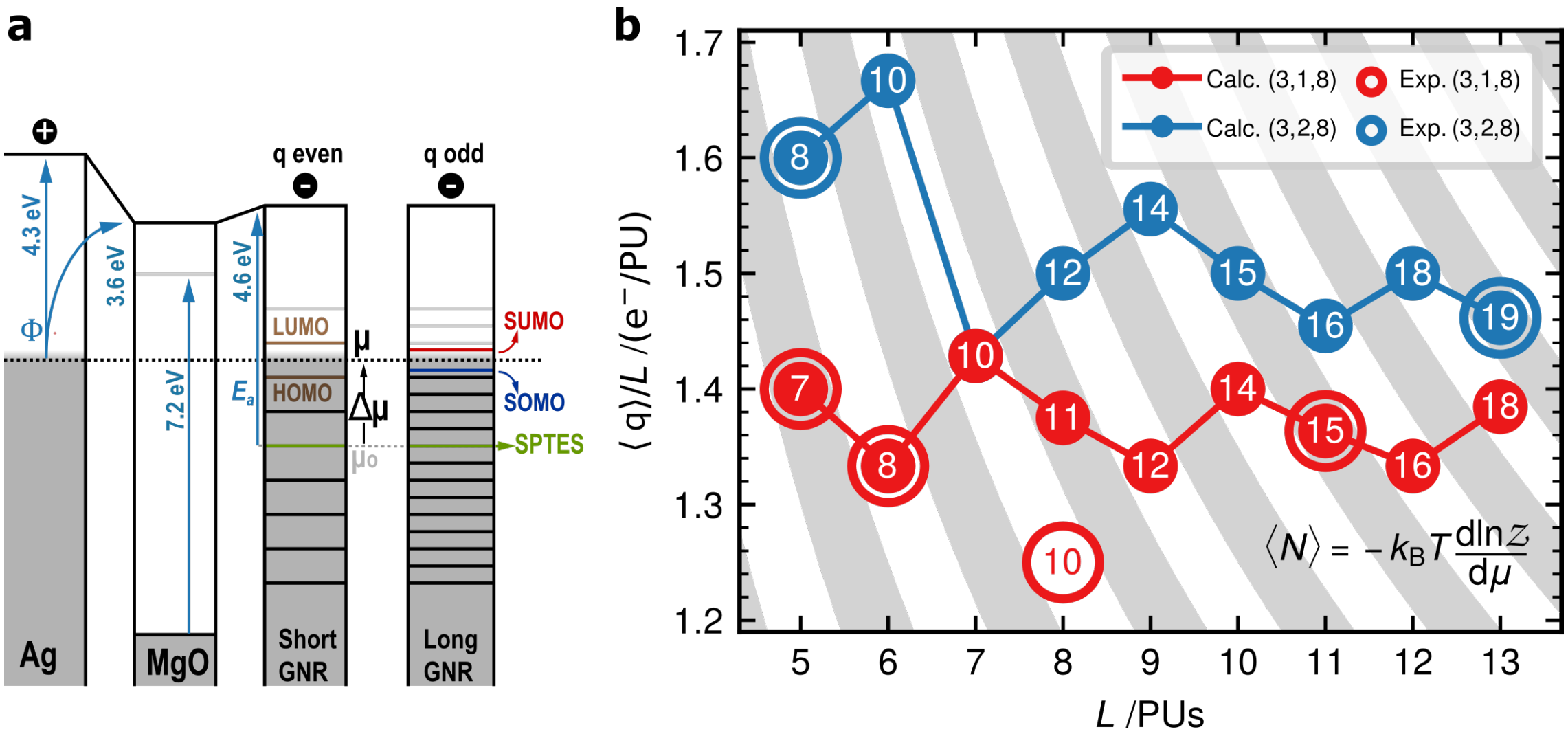

**Figure 5.- Comparison of experimental electron occupation and theoretical calculations. a)** Energy level alignment scheme for short even- and long odd-integer charged GNRs on top of a MgO layer (band gap 7.2 eV [38]), respectively. The neutral level of the GNRs is referenced by the binding energy of the two symmetry-protected topological end states (SPTEs). **b)** Fit of the grand canonical charging model to the experimentally observed charge per precursor unit as a function of ribbon length for the two types of chiral GNRs. Stripes in grey correspond to odd-integer excess charge.

To address this scenario, we developed a theoretical model accounting for electron doping as a function of length and chirality. When molecular species lie on substrates with work function Φ lower than their electron affinity $E_a$, an interfacial charge redistribution takes place, resulting in the electron accumulation into unoccupied molecular states and the build-up a local interface dipole with associated potential energy $U_d$ that opposes to charge transfer [39–41]. In equilibrium, the charge state of the molecule is determined by the shift of $\mu$ (the chemical potential at the metal's Fermi level) from the value at which the molecule retains charge neutrality, $\mu_0$ (Fig. 5a).

$$\Delta\mu = \mu - \mu_0 \; = E_a - \Phi - U_d \quad (1)$$

Since the Ag(001) work function ($\Phi_{Ag(001)}$=4.3 eV [37,40,42]) is smaller than the GNR's electron affinity (comparable to the bulk work function of graphene, $\Phi_{GNR}$~4.61 eV), the nanoribbons exhibit significant n-doping already on Ag(001). As shown in Supplementary Table S1, we find an average doping of 1.25 $e^-$/PU for (3,1,8)-GNR on Ag(100), in close agreement with the previous reported value of 1.3 $e^-$/PU on a similar system [43]. The addition of a MgO insulating layer further reduces the substrate's work function via the pillow effect [37,44,45] by $\delta\Phi$=0.63±0.12 eV (see Supplementary Fig. S7), to yield an exceptionally low work function $\Phi_{Ag/MgO}$~3.7±0.1 eV. The MgO also suppresses the wave-function overlap between molecule and metal such that only integer charges are allowed. Depending on the alignment of $\mu$ with respect to the discrete molecular levels (Fig. 5a), the resulting charge state may be either even or odd. Furthermore, the dielectric spacer reduces the interface capacitance between molecule and metal, and thereby the charge state.

To extract the magnitude and parity of the acquired excess charge ($q$), we represent each GNR in contact with the substrate bath by a chemical potential $\mu$ and temperature $T$ in the grand canonical ensemble. We compute the internal energy of the various charge states of (3,$n$,8)-GNRs ($n$=1,2) with increasing length $L$ using the MFH model (see Supplementary Theoretical Methods) and obtain the mean number of excess electrons $\langle q \rangle$ for any given chemical potential $\mu$ (see Supplementary Fig. S10) via the relation

$$\langle q(\mu, L, T) \rangle = -k_B T \frac{dZ(\mu,L,T)}{d\mu} \qquad (2)$$

where $\mathcal{Z}(\mu, L, T)$ is the grand canonical partition function obtained by summing over all relevant charge states for each GNR (See Supplementary Theoretical Methods).

Our model reproduces that for a fixed chemical potential the excess charge increases with length. In fact, the addition of one additional PU increases $q$ by either zero, one or two electrons, very much in line with the experimental observation of a non-integer mean value of $q/L$. This is due to the intrinsic evolution of the level spacing of the QW edges states of (3,$n$,8)-GNR with length and chirality. In the thermodynamic limit $L \to \infty$, $\langle q \rangle / L$ asymptotically approaches the intensive value corresponding the n-type carrier density of a one dimensional GNR gated by an electric potential $\Delta\mu/|e|$.

To obtain the local gating $\Delta\mu$ of each family of chiral ribbons on the MgO layer, we fit the experimental electron doping deduced in Figs. 3 and 4 with predicted charges using $\Delta\mu$ as a length-independent free parameter. The results of the fit, shown in Fig. 5b, reproduce the charging pattern of (3,1,8)- and (3,2,8)-GNRs with $\Delta\mu$=0.51 eV and 0.49 eV, respectively (see Supplementary Figure S10). The only experimental deviation is the case of (3,1,8)-GNR with $L$=8, for which the model predicts the single occupancy of the 5$^{th}$ QW state. Instead, this GNR appears on MgO charged with ten electrons, with its LUMO (the 5$^{th}$ QW state) at only 17 meV above $E_F$ (Fig. 4a and Supplementary Fig. S8), i.e., at the verge of single occupancy. This is likely due to the electrostatic potential emanating from nearby charged defects in the MgO lattice (see examples in Supplementary Figs. S11 and S12).

The determined $\Delta\mu$ value is about half of the ~0.9 eV work function difference between MgO/Ag(001) and graphene. This can be attributed to the existence of a sizable interface dipole across the MgO layer, which conversely acts to lower the GNR's charge state [41]. From equation (1), the electrostatic energy stored in the interface becomes $U_d$~0.4 eV, which matches well with a plate capacitor model [41] with excess charge lying mainly on the zig-zag edges of the GNRs.

## CONCLUSION

When chiral GNRs are positioned on the MgO monolayer on Ag(001), the combination of its low work function and the electronic decoupling gives rise to quantized charge transfer to their edge states, whose occupation can be controlled by a modification of the length of just one PU. As a consequence, the Fermi level raises up to QW states that lie 300 to 650 meV above in the corresponding charge neutral GNR. Furthermore, the e-e correlations of extended GNR edge states on MgO are sufficiently large as to stabilize singly occupied QW edge states, leading to spin-1/2 quantum dot behaviour at the MFH level of theory, with spin-split frontier states. This is reminiscent of the single electron transistor behaviour of individually contacted GNRs [46], but here the dot occupancy is controlled by the GNR length in addition to a gate voltage. It opens up possibilities for disruptive functional devices such as graphene-based spin qubits and spin-polarized field effect transistors.

Our experimental set up offers an unprecedented combination of ample synthetic capabilities with the utmost sensitivity to the electronic structure. This opens up opportunities to design functional nanographenes with well-defined quantum states, unlike existing strategies to fabricate devices based on GNRs: the transfer of semiconducting GNR arrays to Si wafers, alumina [46] and graphene [47]; the direct synthesis on oxide [25] surfaces, the intercalation of non-metallic layers or functional groups [35]; and the manipulation onto thin layers of NaCl [12,24,26].

For example, on MgO/Ag(001) the energy landscape determined by the electrostatic potential at the tip-sample gap can induce rigid shifts of the whole GNR spectrum of the order of tens of meV (Figs. 2a, 2b and Supplementary Fig. S11), or even switching between odd and even occupancy (Supplementary Figure S12). This can be combined with atomic manipulation of the GNRs, leading to interacting quantum dots with controllable spin and charge states.

## METHODS

**Sample preparation.** Samples are prepared at a base pressure of $1\times10^{-10}$ mbar. The Ag(001) single crystal from SPL B.V. was cleaned by repeated Argon sputtering and annealing at 430 ºC. MgO monolayer patches are grown by depositing Mg from a crucible (MBE Komponenten GmBh effusion cell) heated at 320 ºC onto the clean Ag(001) held at a constant temperature of 390-400 ºC in an O2 partial pressure of $1\times10^{-6}$ mbar. The growth rate of MgO under these conditions fluctuates between 0.5 to 0.1 ML/min. After deposition, we wait a time lapse of 30 min to properly pump down the residual $O_2$ molecules in the chamber ($p<1\times10^{-9}$ mbar), and then we anneal the sample during 20 min at 390 ºC with the purpose of healing the disorder at the edges of the MgO and decrease the number of point defects within the islands (see Supplemental Fig. S1). To synthesize the (3,$n$,1)-GNRs we sublimate onto the Ag(001) surface (already with the MgO patches on it) the precursor reactants **1** and **2** (Fig 1a) that yield GNRs with $n$=1 and $n$=2 respectively and then perform a single annealing step at 345 ºC during 15 min to achieve full cyclodehydrogenation of the GNRs (further details in Supplementary Experimental Methods). The synthesis and quality of the GNRs on Ag(001) with and without coexisting MgO monolayer islands is the same, with the only exception that in the presence of MgO, the GNRs are in average shorter, probably owing to the lower mobility of the shorter precursor oligomers on the surfaces with lower available metallic area.

**Lateral atomic manipulation of GNRs.** GNRs are relocated on the surface by approaching the tip to their arm-chair termini (where large electron density is typically concentrated near Fermi level) until attracting forces between the tip foremost atoms and the ribbon are large enough as to drag the ribbon jumping repeatedly below the tip as it moves laterally. This procedure is illustrated in Supplementary Fig. S3 and S4. To move the GNR across the Ag surface, the tip moves under closed feedback conditions at a tunnelling resistance given by the set point of 3 mV and 10-40 nA. To transfer the a GNR to a MgO patch in close proximity, we use instead open feedback conditions at a similar tunnelling set point of 1-3 mV and 10-60 nA.

**Scanning Tunnelling Microscopy and Spectroscopy.** All measurements have been performed at the SPECS-JT-STM of the Laboratory for Advanced Microscopy (University of Zaragoza). The whole system operates under ultra-high-vacuum conditions ($1\times10^{-10}$ mbar). The tip is grounded and the tunnelling bias $V_b$ is applied to the sample. Data has been taken at $T$=1.2 K unless stated otherwise. Differential tunnelling conductance $dI/dV$ is acquired using a lock-in amplifier at a frequency of 973 Hz and r.m.s. modulation given by $V_{\mathrm{mod}}$. STM images and $dI/dV$ maps were taken either in constant height or in constant current mode, using a stabilization distance determined by the set point indicated at the corresponding caption for each data set. Tips are prepared by electrochemical etching of W wires and subsequent field emission cleaning (120 V, 1 μA, 30 min) at the STM head. CO functionalization of the tip is achieved by controlled approach of the tip to a CO adsorbed on the surface at $|V_b|<5$ mV until a sudden jump in the current is detected.

**Mean-field Hubbard calculations.** The electronic structure and magnetic ground state of GNRs is well captured by the Hubbard model [48]. By employing a mean-field approach [49], we solve for the ground state and obtain eigen-functions and -energies of the $\pi$-electrons, using the same Hamiltonian parametrization as in Refs. [17,19], which has been previously found to reproduce the experimental band gaps in GNRs. We choose a Hubbard parameter of $U$= 3 eV to capture the observed SOMO-SUMO splitting as a function of length and chirality. Differential conductance maps are simulated by squaring a linear combination of $p_z$ orbitals on the lattice at a height of 1.8 nm.

## ACKNOWLEDGEMENTS

The authors gratefully acknowledge financial support from the Spanish MICIU/AEI/10.13039/501100011033 and by "ERDF A way of making Europe" through grants PID2019107338RB-C64, PID2019-107338RB-C61, PID2022-138750NB-C21, PID2022-140845OB-C61, PID2020-115406GB-I00, JDC2022-048665-I and the Excellence Programs Severo Ochoa CEX2023-001286-S and Maria de Maeztu CEX2020-001038-M. This work was also supported by European Regional Development Fund (ERDF) under the program Interreg V-A España-Francia-Andorra (grant no. EFA194/16 TNSI), by the European Union (EU) through the FET-Open project SPRING (863098), the ERC Synergy Grant MolDAM (951519), the ERC-AdG CONSPIRA (101097693), by the Aragon Government (E13-23R and E12-23R), and by the Xunta de Galicia (Centro singular de investigación de Galicia accreditation 2019-2022, ED431G 2019/03 and Oportunius Program).

## REFERENCES

[1] Q. Sun, R. Zhang, J. Qiu, R. Liu, and W. Xu, On-Surface Synthesis of Carbon Nanostructures, Advanced Materials **30**, 1705630 (2018).

[2] S. Clair and D. G. de Oteyza, Controlling a Chemical Coupling Reaction on a Surface: Tools and Strategies for On-Surface Synthesis, Chem. Rev. **119**, 4717 (2019).

[3] J. Cai et al., Atomically precise bottom-up fabrication of graphene nanoribbons, Nature **466**, 470 (2010).

[4] D. G. de Oteyza and T. Frederiksen, Carbon-based nanostructures as a versatile platform for tunable π-magnetism, J. Phys.: Condens. Matter **34**, 443001 (2022).

[5] S. Mishra et al., Topological frustration induces unconventional magnetism in a nanographene, Nat. Nanotechnol. **15**, 22 (2019).

[6] E. Turco, A. Bernhardt, N. Krane, L. Valenta, R. Fasel, M. Juríček, and P. Ruffieux, Observation of the Magnetic Ground State of the Two Smallest Triangular Nanographenes, JACS Au **3**, 1358 (2023).

[7] S. Song et al., Highly entangled polyradical nanographene with coexisting strong correlation and topological frustration, Nat. Chem. **16**, 938 (2024).

[8] J. Hieulle et al., On-Surface Synthesis and Collective Spin Excitations of a Triangulene-Based Nanostar, Angewandte Chemie International Edition **60**, 25224 (2021).

[9] J. Su et al., Atomically precise bottom-up synthesis of π-extended [5]triangulene, Sci. Adv. **5**, eaav7717 (2019).

[10] N. Pavliček, A. Mistry, Z. Majzik, N. Moll, G. Meyer, D. J. Fox, and L. Gross, Synthesis and characterization of triangulene, Nature Nanotech **12**, 308 (2017).

[11] N. Friedrich, R. E. Menchón, I. Pozo, J. Hieulle, A. Vegliante, J. Li, D. Sánchez-Portal, D. Peña, A. Garcia-Lekue, and J. I. Pascual, Addressing Electron Spins Embedded in Metallic Graphene Nanoribbons, ACS Nano **16**, 14819 (2022).

[12] P. Ruffieux et al., On-surface synthesis of graphene nanoribbons with zigzag edge topology, Nature **531**, 489 (2016).

[13] S. Mishra et al., Large magnetic exchange coupling in rhombus-shaped nanographenes with zigzag periphery, Nat. Chem. **13**, 581 (2021).

[14] O. V. Yazyev, R. B. Capaz, and S. G. Louie, Theory of magnetic edge states in chiral graphene nanoribbons, Physical Review B **84**, 115406 (2011).

[15] J. Li, S. Sanz, M. Corso, D. J. Choi, D. Peña, T. Frederiksen, and J. I. Pascual, Single spin localization and manipulation in graphene open-shell nanostructures, Nature Communications **10**, 200 (2019).

[16] S. Mishra, D. Beyer, R. Berger, J. Liu, O. Gröning, J. I. Urgel, K. Müllen, P. Ruffieux, X. Feng, and R. Fasel, Topological Defect-Induced Magnetism in a Nanographene, J. Am. Chem. Soc. **142**, 1147 (2020).

[17] J. Li, S. Sanz, N. Merino-Díez, M. Vilas-Varela, A. Garcia-Lekue, M. Corso, D. G. de Oteyza, T. Frederiksen, D. Peña, and J. I. Pascual, Topological phase transition in chiral graphene nanoribbons: from edge bands to end states, Nat Commun **12**, 5538 (2021).

[18] N. Merino-Díez et al., Unraveling the Electronic Structure of Narrow Atomically Precise Chiral Graphene Nanoribbons, J. Phys. Chem. Lett. **9**, 25 (2018).

[19] J. Brede et al., Detecting the spin-polarization of edge states in graphene nanoribbons, Nat Commun **14**, 6677 (2023).

[20] J. Fischer, B. Trauzettel, and D. Loss, Hyperfine interaction and electron-spin decoherence in graphene and carbon nanotube quantum dots, Phys. Rev. B **80**, 155401 (2009).

[21] J.-S. Chen, K. J. Trerayapiwat, L. Sun, M. D. Krzyaniak, M. R. Wasielewski, T. Rajh, S. Sharifzadeh, and X. Ma, Long-lived electronic spin qubits in single-walled carbon nanotubes, Nat Commun **14**, 848 (2023).

[22] O. Deniz, C. Sánchez-Sánchez, T. Dumslaff, X. Feng, A. Narita, K. Müllen, N. Kharche, V. Meunier, R. Fasel, and P. Ruffieux, Revealing the Electronic Structure of Silicon

Intercalated Armchair Graphene Nanoribbons by Scanning Tunneling Spectroscopy, Nano Lett. **17**, 2197 (2017).
[23] A. Kinikar et al., Electronic Decoupling and Hole-Doping of Graphene Nanoribbons on Metal Substrates by Chloride Intercalation, ACS Nano **18**, 16622 (2024).
[24] S. Wang, L. Talirz, C. A. Pignedoli, X. Feng, K. Müllen, R. Fasel, and P. Ruffieux, Giant edge state splitting at atomically precise graphene zigzag edges, Nature Communications **7**, 11507 (2016).
[25] M. Kolmer, A.-K. Steiner, I. Izydorczyk, W. Ko, M. Engelund, M. Szymonski, A.-P. Li, and K. Amsharov, Rational synthesis of atomically precise graphene nanoribbons directly on metal oxide surfaces, Science **369**, 571 (2020).
[26] P. H. Jacobse, M. J. J. Mangnus, S. J. M. Zevenhuizen, and I. Swart, Mapping the Conductance of Electronically Decoupled Graphene Nanoribbons, ACS Nano **12**, 7048 (2018).
[27] J. Repp, G. Meyer, S. Stojković, A. Gourdon, and C. Joachim, Molecules on Insulating Films: Scanning-Tunneling Microscopy Imaging of Individual Molecular Orbitals, Physical Review Letters **94**, (2005).
[28] J. Martinez-Castro, M. Piantek, S. Schubert, M. Persson, D. Serrate, and C. F. Hirjibehedin, Electric polarization switching in an atomically thin binary rock salt structure, Nature Nanotechnology **13**, 19 (2018).
[29] W. Steurer, L. Gross, and G. Meyer, Local thickness determination of thin insulator films via localized states, Appl. Phys. Lett. **104**, 231606 (2014).
[30] F. D. Natterer, K. Yang, W. Paul, P. Willke, T. Choi, T. Greber, A. J. Heinrich, and C. P. Lutz, Reading and writing single-atom magnets, Nature **543**, 226 (2017).
[31] K. Yang, W. Paul, S.-H. Phark, P. Willke, Y. Bae, T. Choi, T. Esat, A. Ardavan, A. J. Heinrich, and C. P. Lutz, Coherent spin manipulation of individual atoms on a surface, Science **366**, 509 (2019).
[32] S. Kovarik, R. Schlitz, A. Vishwakarma, D. Ruckert, P. Gambardella, and S. Stepanow, Spin torque–driven electron paramagnetic resonance of a single spin in a pentacene molecule, Science **384**, 1368 (2024).
[33] G. Reecht, N. Krane, C. Lotze, L. Zhang, A. L. Briseno, and K. J. Franke, Vibrational Excitation Mechanism in Tunneling Spectroscopy beyond the Franck-Condon Model, Phys. Rev. Lett. **124**, 116804 (2020).
[34] J. van der Lit, M. P. Boneschanscher, D. Vanmaekelbergh, M. Ijäs, A. Uppstu, M. Ervasti, A. Harju, P. Liljeroth, and I. Swart, Suppression of electron–vibron coupling in graphene nanoribbons contacted via a single atom, Nat Commun **4**, 2023 (2013).
[35] M. S. G. Mohammed, L. Colazzo, R. Robles, R. Dorel, A. M. Echavarren, N. Lorente, and D. G. de Oteyza, Electronic decoupling of polyacenes from the underlying metal substrate by sp3 carbon atoms, Commun Phys **3**, 159 (2020).
[36] M. Bieletzki, T. Hynninen, T. M. Soini, M. Pivetta, C. R. Henry, A. S. Foster, F. Esch, C. Barth, and U. Heiz, Topography and work function measurements of thin MgO(001) films on Ag(001) by nc-AFM and KPFM, Phys. Chem. Chem. Phys. **12**, 3203 (2010).
[37] M. Hollerer, D. Lüftner, P. Hurdax, T. Ules, S. Soubatch, F. S. Tautz, G. Koller, P. Puschnig, M. Sterrer, and M. G. Ramsey, Charge Transfer and Orbital Level Alignment at Inorganic/Organic Interfaces: The Role of Dielectric Interlayers, ACS Nano **11**, 6252 (2017).
[38] S. Schintke, S. Messerli, M. Pivetta, F. Patthey, L. Libioulle, M. Stengel, A. De Vita, and W.-D. Schneider, Insulator at the Ultrathin Limit: MgO on Ag(001), Phys. Rev. Lett. **87**, 276801 (2001).
[39] M. Fahlman, S. Fabiano, V. Gueskine, D. Simon, M. Berggren, and X. Crispin, Interfaces in organic electronics, Nat Rev Mater **4**, 627 (2019).
[40] M. Willenbockel, D. Lüftner, B. Stadtmüller, G. Koller, C. Kumpf, S. Soubatch, P. Puschnig, M. G. Ramsey, and F. S. Tautz, The interplay between interface structure, energy level alignment and chemical bonding strength at organic–metal interfaces, Phys. Chem. Chem. Phys. **17**, 1530 (2015).

[41] P. Hurdax, M. Hollerer, P. Puschnig, D. Lüftner, L. Egger, M. G. Ramsey, and M. Sterrer, Controlling the Charge Transfer across Thin Dielectric Interlayers, Adv. Mater. Inter. **7**, 2000592 (2020).
[42] G. N. Derry, M. E. Kern, and E. H. Worth, Recommended values of clean metal surface work functions, J. of Vac. Sci. & Tech. A **33**, 060801 (2015).
[43] M. Corso, R. E. Menchón, I. Piquero-Zulaica, M. Vilas-Varela, J. E. Ortega, D. Peña, A. Garcia-Lekue, and D. G. de Oteyza, Band Structure and Energy Level Alignment of Chiral Graphene Nanoribbons on Silver Surfaces, Nanomaterials **11**, 3303 (2021).
[44] G. Witte, S. Lukas, P. S. Bagus, and C. Wöll, Vacuum level alignment at organic/metal junctions: "Cushion" effect and the interface dipole, Appl. Phys. Lett. **87**, 263502 (2005).
[45] H.-J. Freund and G. Pacchioni, Oxide ultra-thin films on metals: new materials for the design of supported metal catalysts, Chem. Soc. Rev. **37**, 2224 (2008).
[46] J. Zhang et al., Contacting individual graphene nanoribbons using carbon nanotube electrodes, Nat Electron **6**, 572 (2023).
[47] M. El Abbassi et al., Controlled Quantum Dot Formation in Atomically Engineered Graphene Nanoribbon Field-Effect Transistors, ACS Nano **14**, 5754 (2020).
[48] J. Hubbard, Electron correlations in narrow energy bands III. An improved solution, Proc. R. Soc. Lond. A **281**, 401 (1964).
[49] S. Sanz, N. Papior, G. Giedke, D. Sánchez-Portal, M. Brandbyge, and T. Frederiksen, Spin-Polarizing Electron Beam Splitter from Crossed Graphene Nanoribbons, Phys. Rev. Lett. **129**, 037701 (2022).

## EXTENDED DATA

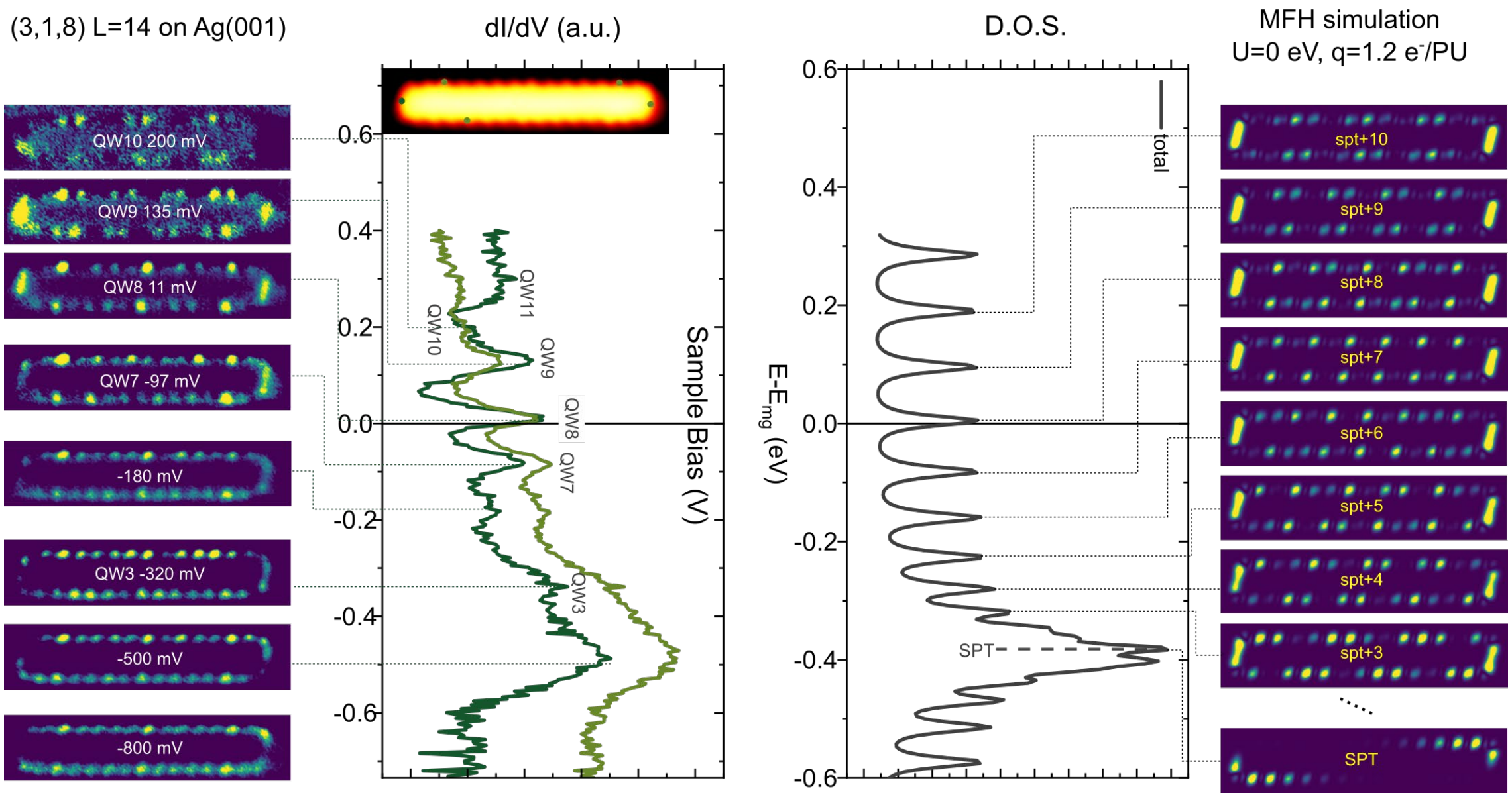

**Extended Data Figure 1.-** From left to right, constant current dI/dV maps of several quantum-well (QW) states of the 14 P.U. (3,1,8)-GNR on Ag(001), corresponding dI/dV spectra, calculated total density of states (DOS) within MFH model (see Methods) and the corresponding simulation of the spatial distribution of each eigenstate. Experimental dI/dV spectra are taken in the positions marked by the dots with the same colour code. STM parameters for maps and spectra: regulation set point 0.5 V/200 pA, $V_{mod}$=5 mV. The broad peak found at around -0.5 V can be explained as the overlap of several low order QW resonances next to the SPT state of the charge neutral GNR. MFH simulations are obtained for a negative charge excess equivalent to16.8 electrons and U=0 eV. DOS is schematically represented by the eigenenergies with a Lorentzian broadening of 20 meV, and simulations of the DOS maps are taken at a height of 1.8 nm.

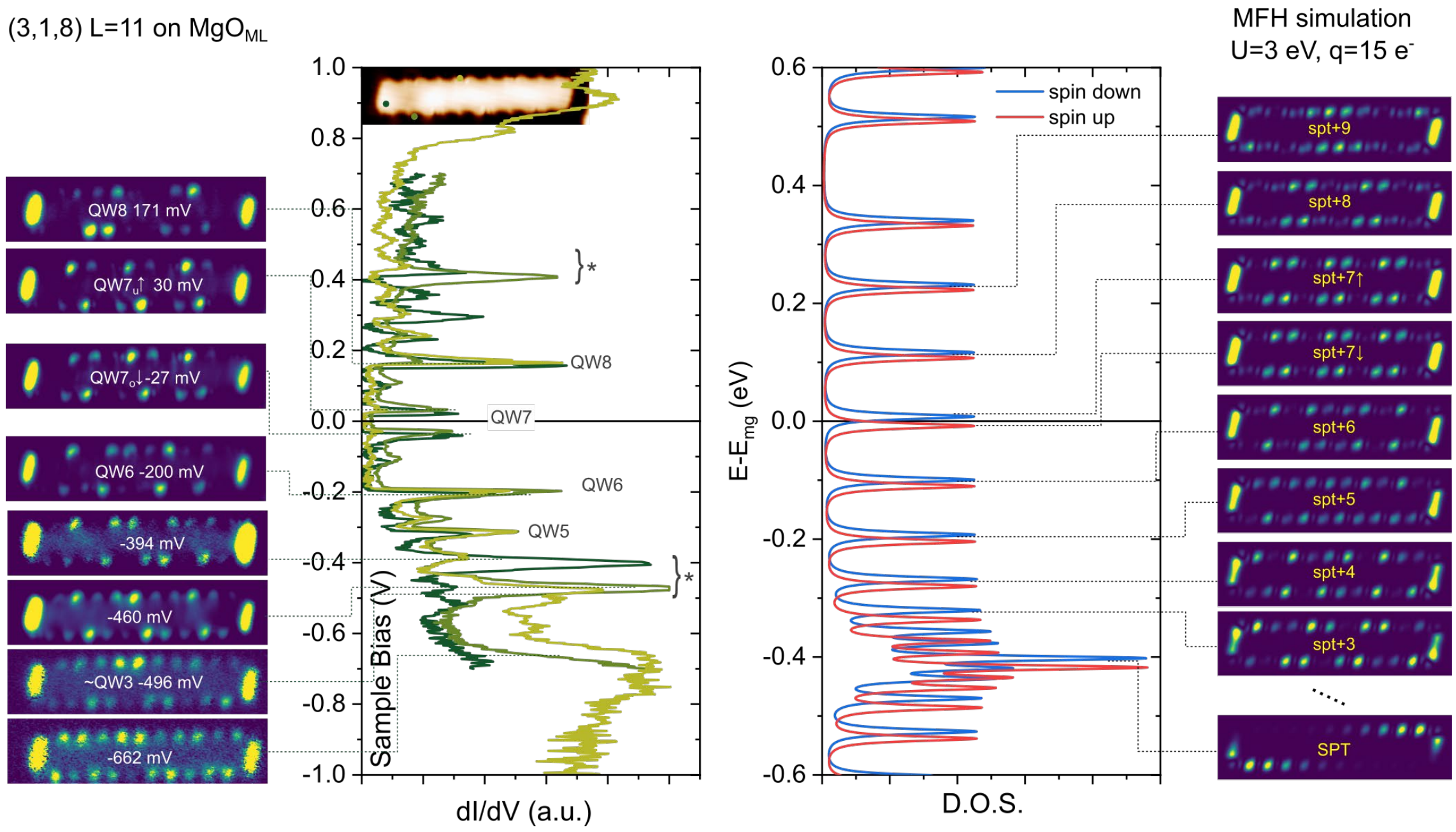

**Extended Data Figure 2.-** From left to right, constant height dI/dV maps of several quantum-well (QW) states of the 11 P.U. (3,1,8)-GNR on MgO, corresponding dI/dV spectra, calculated total density of states (DOS) within MFH model (see Methods) and theoretical simulation of the spatial distribution of each eigenstate. Experimental dI/dV spectra are taken in the positions marked by the dots with the same colour code. STM parameters for maps: regulation set point at ribbon centre is 50 mV/20[100/5] pA for QW8[QW7/QW6] and 0.5V/200 pA for the rest, $V_{\text{mod}}$=1 mV. STM parameters for spectra: stabilization at -0.5 V/200 pA, $V_{mod}$=2 mV. The images of occupied/unoccupied spin split QW7 states are labelled by down/up arrows. The charging peaks are indicated by an asterisk (*). The broad peak found at around -0.65 eV can be explained as the overlap of several low order QW resonances next to the SPT state of the charge neutral GNR. MFH simulations are obtained for 15 electrons excess and U=3 eV. DOS is schematically represented by the eigenenergies with a Lorentzian broadening of 2 meV, and simulations of the DOS maps are taken at a height of 1.8 nm.

Supplementary Information for:

# Systematic Modulation of Charge and Spin in Graphene Nanoribbons on MgO

Amelia Domínguez-Celorrio[1,2,3,†], Leonard Edens[4, †], Sofía Sanz[5], Manuel Vilas-Varela[6], Jose Martinez-Castro[7], Diego Peña[6], Véronique Langlais[8], Thomas Frederiksen[5,9], José I. Pascual[4,9], and David Serrate[1,10,11*]

[1]Insituto de Nanociencia y Materiales de Aragón (INMA), CSIC-Universidad de Zaragoza, Zaragoza, E-50009, Spain

[2]School of Physics and Astronomy, Monash University, Clayton, VIC 3800, Australia

[3]ARC Centre for Future Low Energy Electronics Technologies, Monash University, Clayton, VIC 3800, Australia

[4]CIC NanoGUNE BRTA, San Sebastián, E-20018, Spain

[5]Donostia International Physics Center, San Sebastián, E-20018, Spain.

[6]Centro Singular de Investigación en Química Bilóxica e Materiais Moleculares (CiQUS) and Departamento de Química Orgánica, Universidade de Santiago de Compostela, Santiago de Compostela, E-15782, Spain.

[7]Peter Grünberg Institut (PGI-3), Forschungszentrum Jülich, 52425 Jülich, Germany

[8]Centre d'Elaboration de Materiaux et d'Etudes Structurales, CNRS, Toulouse, F-31055 France

[9]Ikerbasque, Basque Foundation for Science, Bilbao, E-48013, Spain.

[10]Departamento de Física de la Materia Condensada, Universidad de Zaragoza, Zaragoza, E-50009, Spain

[11]Laboratorio de Microscopias Avanzadas (LMA), Universidad de Zaragoza, Zaragoza, E-50018, Spain.

†These authors contributed equally

*email: serrate@unizar.es

## Table of Contents:

# I. Supplementary Experimental Methods

MgO ultrathin patches on Ag(001) are grown by depositing Mg from an alumina crucible) heated at 320-330 ºC onto the clean Ag(001) held at a constant temperature of 390-400 ºC in an O2 partial pressure of $1\times10^{-6}$ mbar. The growth rate of MgO under these conditions fluctuates between 0.5 to 0.1 ML/min. After deposition, we wait a time lapse of 30 min to properly pump down the residual $O_2$ molecules in the chamber ($p<1\times10^{-9}$ mbar), and then we anneal the sample during 15-20 min at 390 ºC with the purpose of healing the disorder at the edges of the MgO and decrease the number of point defects within the islands (see Fig. S1).

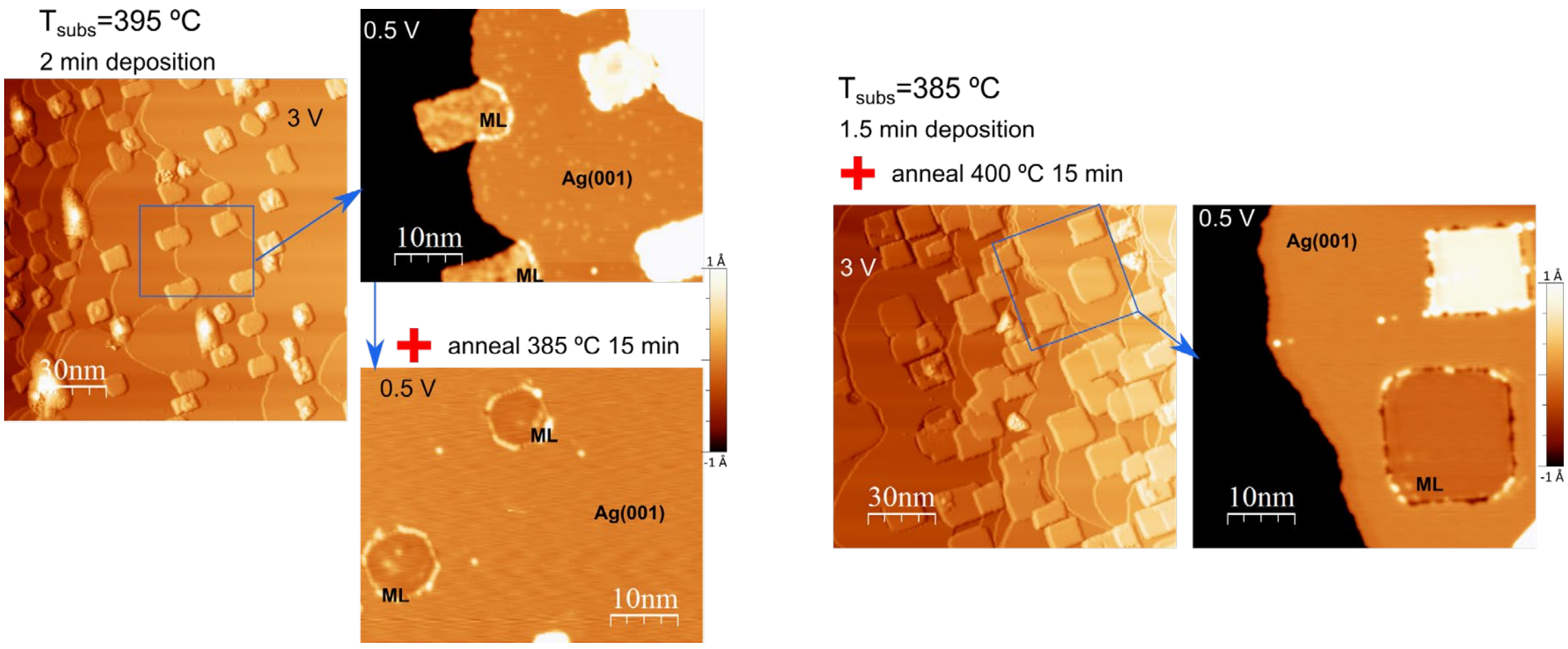

**Figure S1.-** Growth parameters of MgO ultrathin islands on Ag(001). STM images (150x150 nm) obtained at a sample bias of 3 V are overviews of two samples grown under similar conditions (fake colour scale to better appreciate the island roughness). Left panel correspond to the as-grown sample while the right panel corresponds to a sample which was post-annealed a 400ºC. The zoomed regions are imaged at 0.5 V. This bias allows us to distinguish between embedded monolayer islands (labelled as ML) and other thicknesses. The ML islands become nearly defect free after post-annealing under UHV conditions.

(3,*n*,1)-GNRs synthesis starts by sublimating precursors **1** and **2** (Figure S1a and S1e) onto the Ag(001) surface where MgO monolayer ($MgO_{ML}$) patches have been previously grown, for *n*=1 and *n*=2 respectively. Ullmann coupling of precursor **1** takes place at room temperature as they reach the Ag(001) surface. As shown in Fig. S1b, the precursor units assemble as short oligomers, mostly attached to the edges of Ag terraces. Upon annealing to 300 ºC (Figure S1c), we distinguish a clear change in the morphology of the sample with longer and ordered chains in islands. The distance between the protrusions shown in Figure S1c is 8.7(1) Å, which corresponds with the expected periodicity of poly-(**1**) chains. Mild annealing processes of poly-(**1**) samples above 300 ºC led to partial cyclodehydrogenation (CDH) of the polymeric chains (Fig. S1c). Fully planarized (3,1,8)-GNRs form after 15 minutes at 345 ºC (Fig. S1d).

Similarly, as illustrated by Fig. S1e, precursor **2** also undergoes Ullmann coupling on Ag(001) at room temperature. In this case, densely packed and ordered poly-(**2**) chains appear without any additional annealing for sufficiently large coverages (as in the left side terrace of Fig. S1f). The characteristic period of these chains is 10.9(1) Å, which corresponds with the expected periodicity of poly-(**2**) chains (Fig. S1g). Annealing of poly-(**2**) samples at 345 ºC during 15 minutes led, as in the case of precursor **1**, to a complete CDH and the formation of (3,2,8)-GNRs (Fig. S1h).

The synthesis and quality of the GNRs on Ag(001) with and without coexisting $MgO_{ML}$ islands is the same, with the only exception that in the presence of MgO, the GNRs are in average shorter, probably owing to the lower mobility of the shorter precursor oligomers on the surfaces with lower available metallic area.

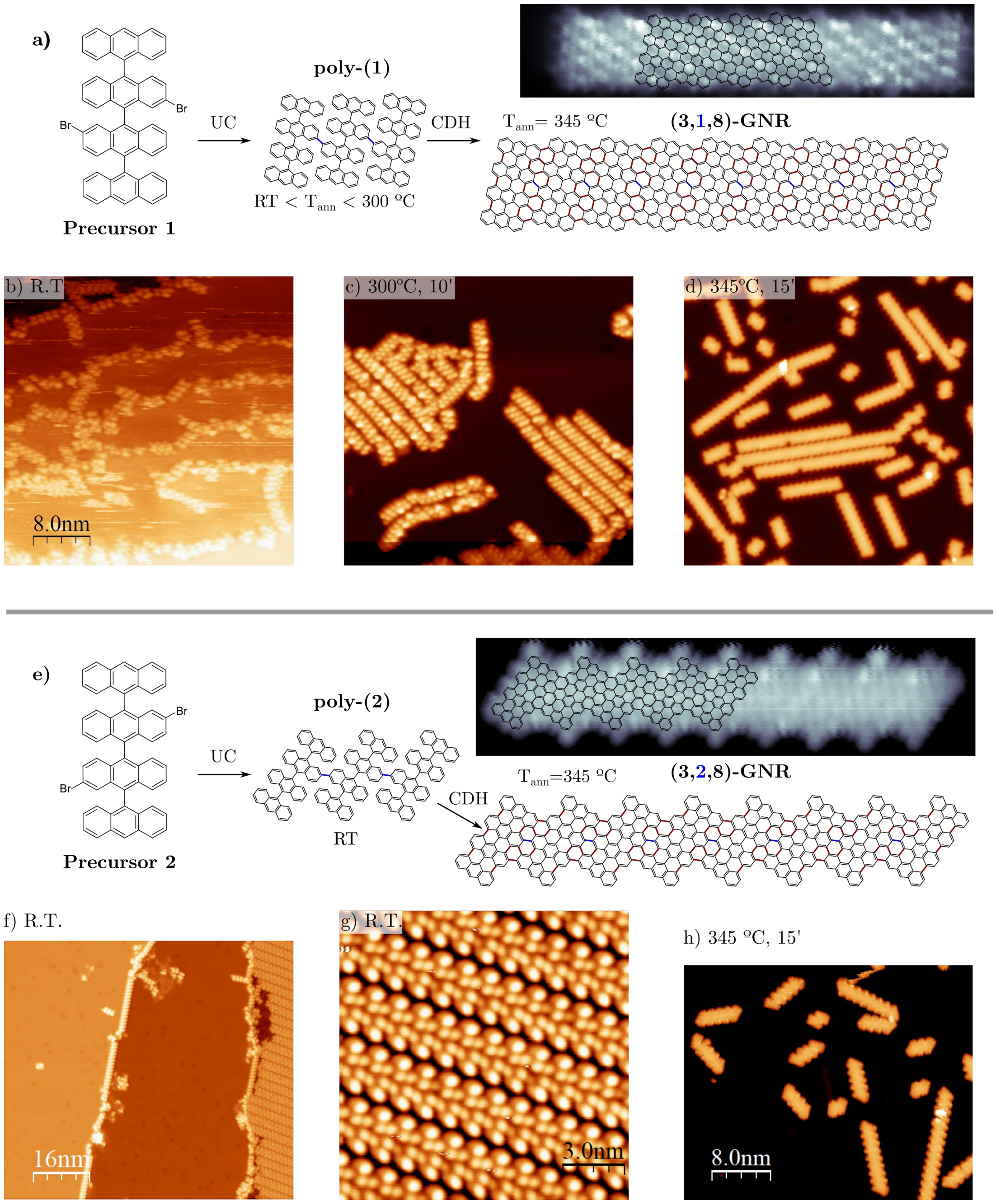

**Figure S2**.- On-surface synthesis of (3,*n*,8)-GNRs on Ag(001) for *n*=1 (a-d) and *n*=2 (e-h). Panels (a) and (e) show the reaction schemes of precursors **1** and **2** respectively showing that Ullmann polymerization takes place already at room temperature, and that short annealing of the polymeric chains at 345 ºC gives rise to a complete cyclodehydrogenation (CDH) in both cases. Regulation $V_b$ of STM topographies is 0.5 V in panels (b-d) , 1.8 V in panel (f), 0.05 V in panel (g), and 0.5 V in panel (h). High resolution insets in (a) and (e) are constant height tunnelling current images with functionalized tips taken at 1.5 mV and 500 mV for (3,1,8)- and (3,2,8)-GNRs respectively.

The direct growth on MgO surfaces is completely discarded, because the aromatic precursors do not adsorb on MgO at room temperature or higher (heating is absolutely necessary to induce dehalogenation). Furthermore, we are not aware of dry or wet transfer stamp techniques from catalytic surfaces to MgO films.

We have also explored the MgO intercalation after the growth of (3,1,8)-GNRs on clean Ag(001). We deposited directly MgO at $p_{O2}=1\times10^{-6}$ mbar with the GNR/Ag(001) held at 220 ºC, a temperature low enough as to preserve the GNRs integrity, and sufficiently high as to form $MgO_{ML}$. The edes of the GNR become decorated with multiple defects. The strategy of depositing Mg and post oxidizing it in a second step does not work either. Metallic Mg intercalates below the GNRs forming a kind of alloyed surface. However, subsequent exposure to $O_2$ results again in defective GNR edges, as found previously for 5-AGNRs [1].

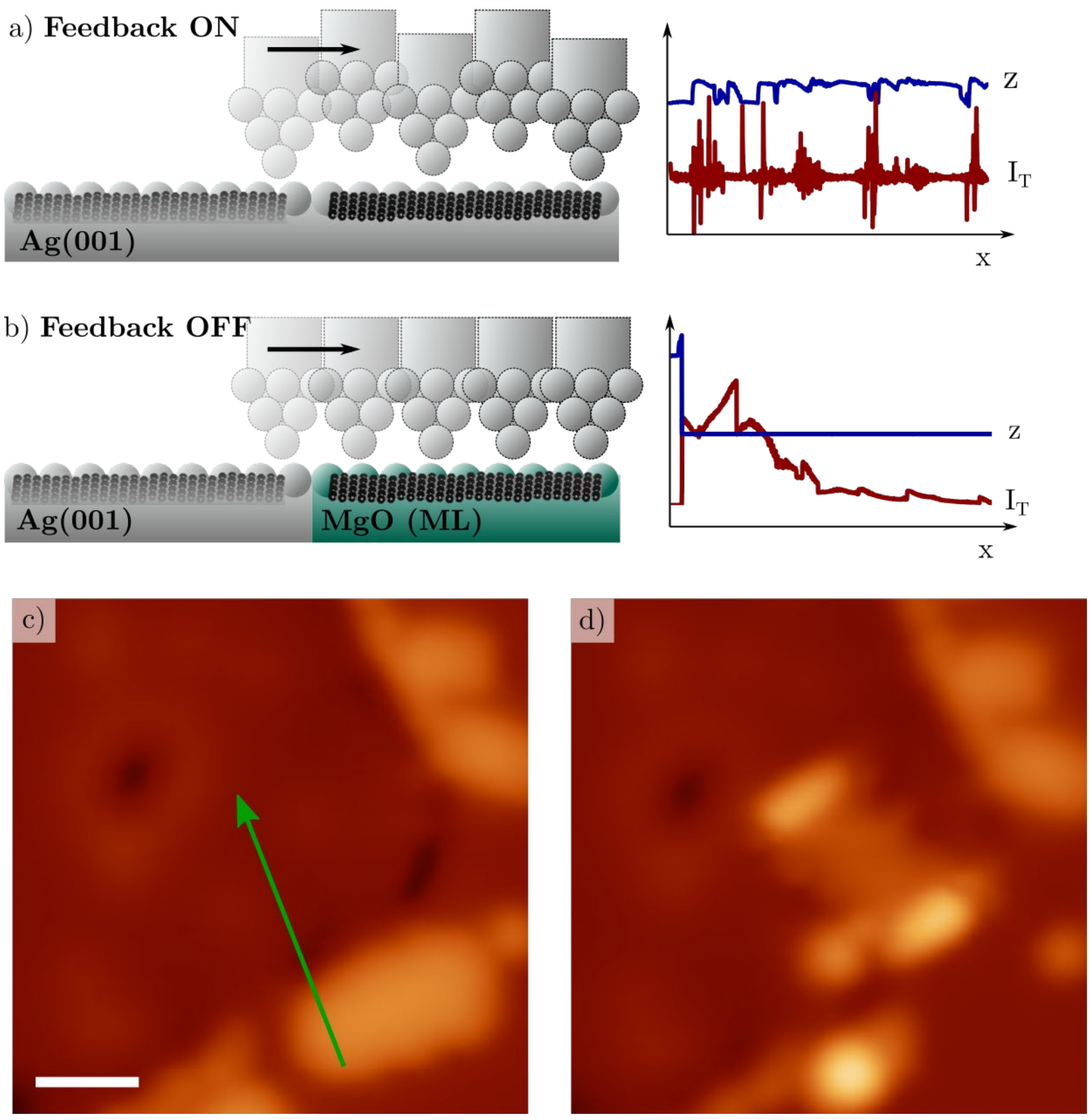

**Figure S3.-** Lateral atomic manipulation of GNRs in pulling mode. (a) Sketch and manipulation traces of GNR on Ag(001). The tip exerts an attracting force over the termini of a GNR and moves laterally in constant current mode (feedback on). $I_T$ is the tunnelling current and z the vertical position of the tip. (b) Sketch and manipulation traces of the transfer of a GNR from across the Ag-MgO boundary. The tip exerts an attracting force over the termini of a GNR and moves in constant height mode (feedback off) to avoid the abrupt jump down that would occur if the feedback was active. (c-d) Example of the successful transfer of a $L$=3 (3,1,8)-GNR. The traces in (a) and (b) correspond to exactly this example, which happened to be the first successful transfer in our project.

Therefore, we decided to resort to the relocation of the GNR by lateral atomic manipulation. Fig. S3 shows schematically the manipulation procedure, together with manipulation traces characteristic of relocations in the Ag surface, and the transfer of the GNR to the MgO island across the MgO-Ag boundary. Lateral manipulation is performed in pulling regime with a gap resistance of the order of 10-50 nA at 3-5 mV bias (sketch of the process in Supplementary Fig. S3a). The preferred contact points to trap the ribbon with the tip force field are the arm-chair termini, which is in line with the large density of states observed at this position (all QW states of the discretized conduction band display strog intensity at the termini, see for instance Supplementary Fig. S8). In average, the manipulation throughout the Ag surface is relatively simple and comparable with the case of individual 3d atoms on coinage metals. On the contrary, the success rate of the transfer from Ag to a MgO patch is much lower (sketch of the process in Supplementary Fig. S3b), and has to be performed under constant height conditions.

## II. Supplementary spectroscopy data.

Figure S4 shows, for (3,1,8)-GNRs on Ag(001), the gradual closing of the gap between the first fully unoccupied quantum well (QW) state and the immediately preceding in energy QW state. As the energy spacing between adjacent QW states decreases for increasing length *L*, the associated *dI/dV* resonances become slightly narrower. STM images on the left side of Fig. S4 illustrate the distinct change in apparent shape of the GNRs as they are stepwise introduced onto the $MgO_{ML}$ patch.

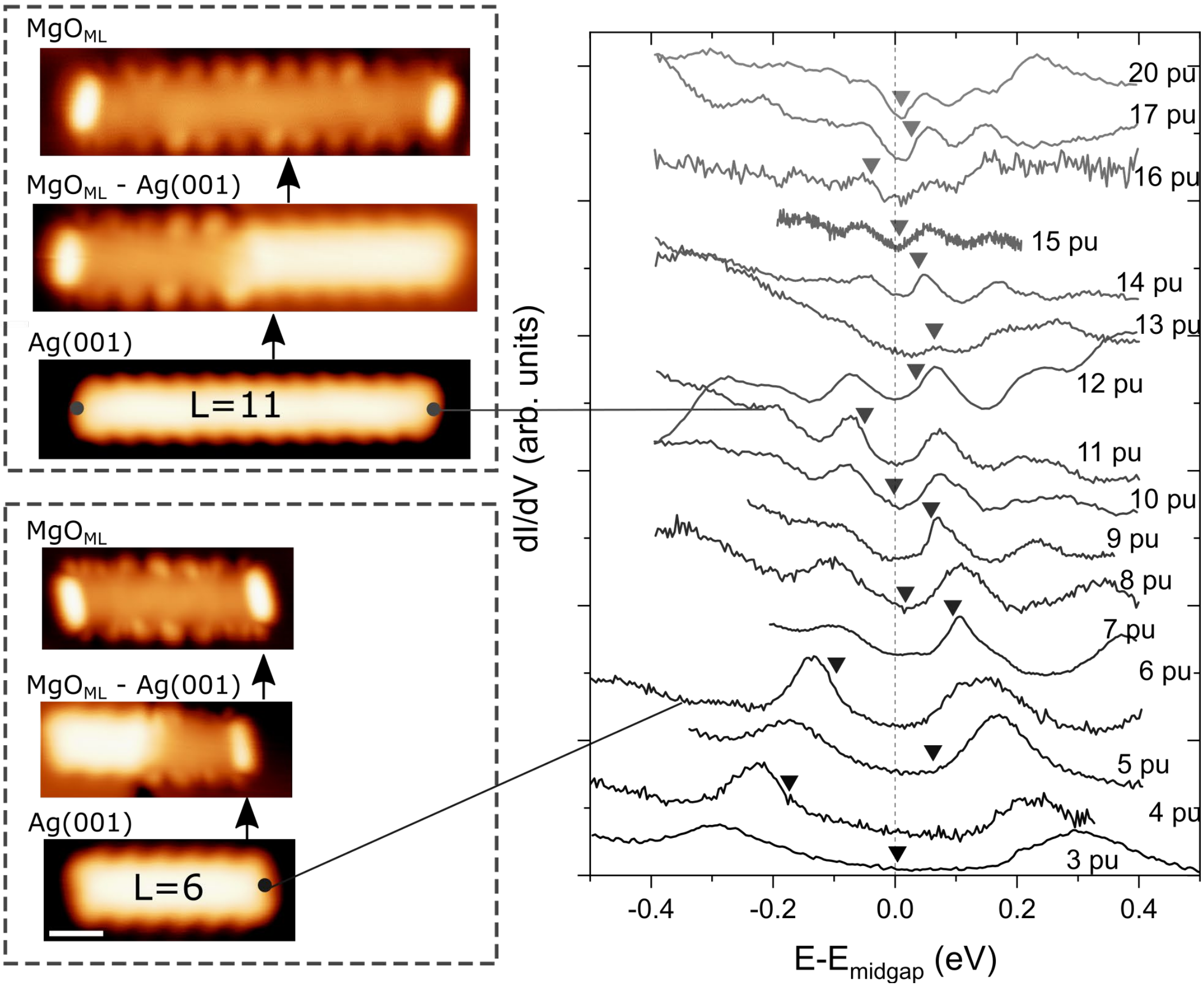

**Figure S4.-** *dI/dV* spectra of (3,1,8)-GNRs on Ag001 for several lengths. Topography and spectroscopy stabilization set points are 0.5 V and 50 pA. Lock-in modulation 5 mV r.m.s. Spectra have been vertically offseted and normalized for the sake of clarity. In the abscissa axis we represent the energy difference with respect to the midgap energy, $E_{\mathrm{midgap}}$. rather than the traditional convention of referring energy with respect to Fermi level. $E_{\mathrm{midgap}}$ is defined as the middle point between the central energy of the first unoccupied (or partially unoccupied) state and that of the immediately preceding state in energy. The experimental Fermi level in each spectrum is indicated by triangles. In this way, the evolution of the gap between QW states nearest to Fermi level can be better appreciated. The left column shows topography images of (3,1,8)-GNRs on Ag(001) with *L*=6 and 11 and, subsequently in the vertical direction, the same GNRs after lateral manipulation: partially and fully inserted in the $MgO_{ML}$. The black circles mark the position where the corresponding *dI/dV* spectra were acquired. The spectra of the other ribbons were also taken at the arm-chair termini, where all QW states feature some intensity. Nevertheless, QW states at the same set of energies can be detected along the chiral edge, though they are often much less intense and not all of them manifest in the same position as a consequence of their intrinsic intensity pattern.

One of the most striking changes observed on the MgO is the extremely low linewidth of the QW resonances. We have studied their full width at half maximum (FWHM) and the results are presented in Fig. S5. At the lowest experimental temperature $T$=1.13 K the linewidth of the 3rd QW state of $L$=5 (3,1,8)-GNR decreases down to 1.3 mV as the lock-in modulation amplitude is reduced stepwise to values smaller than $k_B T/|e|$~0.1 mV.

A)

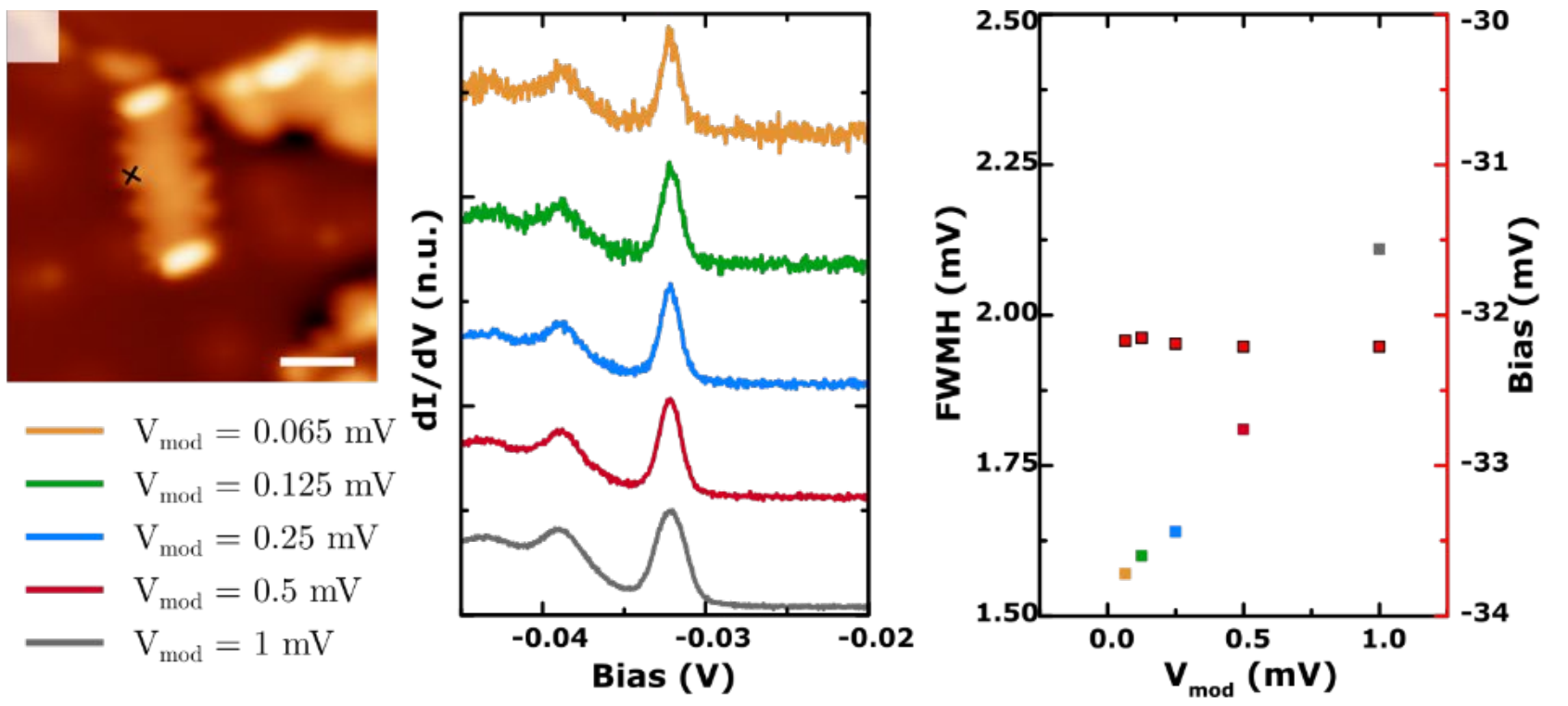

**Figure S5.-** STM topography image of a $L$ = 5 (3,1,8)-GNRs on $MgO_{ML}$ (0.5 V, 200 pA, $T$ = 1.1 K, scale bar = 2 nm) and *dI/dV* point spectra recorded at the position of the cross with varying $V_{mod}$ between 1 and 0.065 mV r.m.s. The left panel shows the experimentally determined FWHM (left axis, colour coded squares) and bias peak values (right axis, red squares) as a function of $V_{mod}$.

These sharp resonances are always accompanied by broader peaks that appear at 7.5 mV and 76 mV (see Fig. S6a-c). This separation is constant throughout the GNR and independent of its length. Constant height *dI/dV* maps of the principal resonance (P, the sharp one) and their replicas (R) unveil that they have identical spatial distribution (Fig. S6a). This overall behaviour is characteristic of vibrational Franc Condon resonances excited as a consequence of long lived ionic GNR states as electrons or holes are injected into the QW states during the tunnelling process [2–4].

We have performed atomistic simulations the vibrational spectra of GNRs on MgO monolayer using harmonic forcefields (see computational details and analysis in Supplementary Theoretical Methods section IV.2). Our analysis suggests that the satellite peaks experimentally observed in dI/dV and labelled as $R_1$ and $R_2$ are, respectively, ascribed to the excitation of the external out-of-plane vibration of whole GNR, and to the out-of-plane H motion. In spite the external origin of the FC resonances, the associated modes happen to be nearly independent of the GNR length, in good agreement with the calculations.

As shown in Figure S6b-c, the vibrational excitations thresholds of $R_1$and $R_2$ are the same for occupied and unoccupied states, and are also the same for QW resonances of different order. They also keep the same values irrespectively of the open or closed shell state of the GNRs (cf. for instance with the example of the L=5 GNR in Fig. S6). This indicates that these vibrations are an intrinsic property of the GNR geometry and adhesion to the MgO, do not depend on the charge or spin states, and can be excited by tunnelling electrons to/from any molecular resonance.

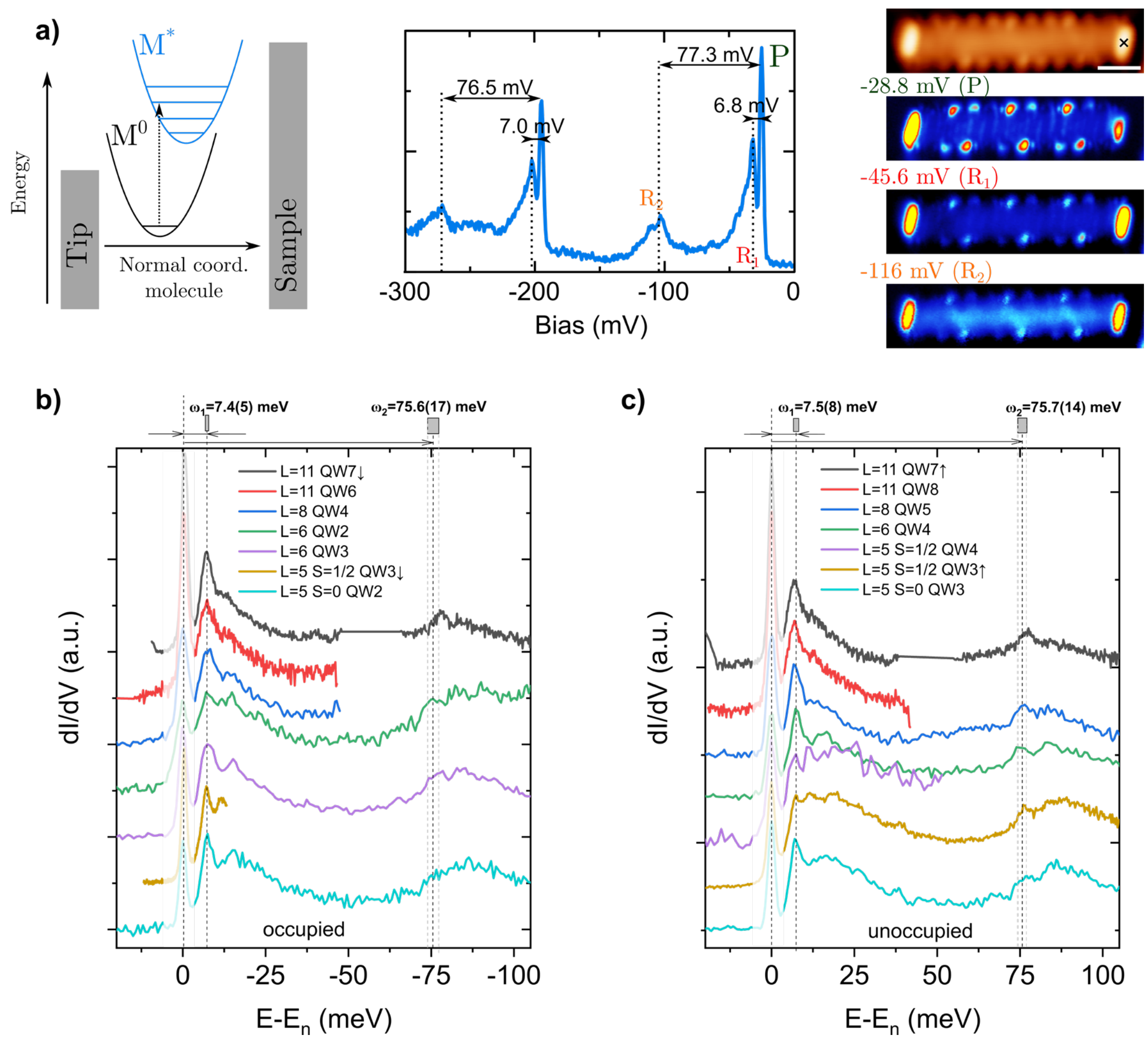

**Figure S6.** Frank-Condon (FC) resonances and theoretical vibrational density of states. **(a)** Sketch of Franck-Condon (FC) mechanism in an STM junction (adapted from ref. [2]) and example of *dI/dV* spectra with satellite resonances recorded for the (3,1,8)-GNR with $L$=11. The FC resonances are labelled as P, for the main peak, and $R_1$ and $R_2$ for the satellite peaks. $V_{mod}$=1 mV r.m.s. The right panel displays the topography of the GNR (the cross indicates the position for spectroscopy) and the constant height *dI/dV* maps recorded at the energies of the main and satellite peaks observed. Notice that all three peaks exhibit identical spatial distribution. Set point for topography, *dI/dV* stabilization and feedback opening in *dI/dV* maps is 0.5 V and 200 pA. Lock-in modulation is 1 mV r.m.s. for the *dI/dV* spectrum and 4 mV r.m.s. for the maps. **(b,c)** Selected high resolution spectra of QWn resonances and accompanying vibrations FC satellites for different energies, GNR lengths and singly/doubly occupancy of frontier states. The peak corresponding to the molecular state, and replicas $R_1$ and $R_2$ are indicated by dotted lines. The abscissae are the energy difference with respect to the corresponding peak energy of each QWn state. Note that (c) shows occupied states and so the energies are negative, whereas (d) shows unoccupied states. Singly occupied/unoccupied molecular states are represented by down/up arrows. In order to facilitate the comparison with the theoretical panel the main molecular state is shaded. $\omega_{1,2}$ stand for the energy of the most intense vibrational modes (R1,2 in panel a) obtained as the average value extracted from 14 different spectra for each bias sign with different QW order (n) and GNR legth (L). Lock-in modulation: 0.2 mV rms in the spectra of $L$=11, 8 and 5-QW3↓, 1 mV rms for the rest.

Upon insertion of the GNRs onto the MgO, we observe a clear trend of the QW states to shift to lower energies. This is well understood if we look at the variation of the local work function difference ($\delta\Phi_{[Ag\text{-}MgO]}$ in the main text) between the bare Ag(001) and the embedded $MgO_{ML}$/Ag(001). To characterize this difference, we acquired field emission resonances inside and outside of the MgO island shown in Fig. S7a. We observe a marked decrease in the bias difference between consecutive resonances (Fig S7b) in MgO, characteristic of a much lower workfunction. Applying a parallel plate capacitor model for the tip-sample gap [5,6], we can use the higher order resonances to give an accurate estimate of $\delta\Phi$=0.63±0.12 eV following the analysis shown in Fig. S7d.

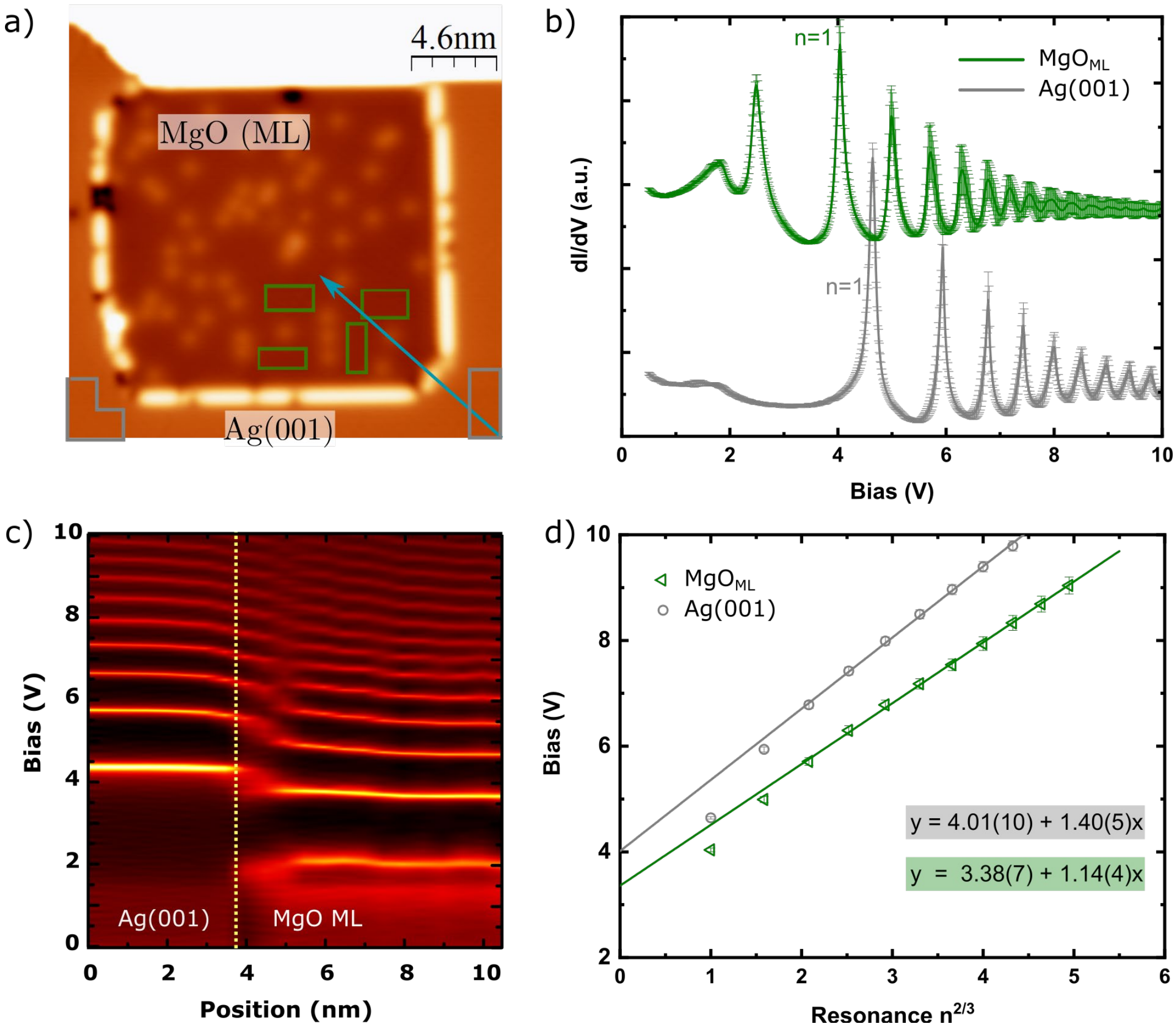

**Figure S7**.- a) STM image of a $MgO_{ML}$ island embedded in Ag(001) (0.5V and $I_t$=100 pA). b) Field emission resonance (FER, stabilization at 0.5 V and 100 pA) measurements recorded on Ag(001) and $MgO_{ML}$ inside the regions enclosed by the rectangles. Solid lines are mean values and error bars are the standard deviation of a set of 90 and 66 curves for Ag and MgO respectively. c) Stack plot of *dI/dV* spectra recorded along the blue arrow in a). The dotted yellow line corresponds to the Ag-MgO intersection in STM images. d) Peak voltages of FER shown in b) plotted against the resonance order to the power of 2/3. The intersect of the linear fits with the ordinate axis gives a very precise estimate of the work function values (the two first FERs are disregarded because in this regime the 1D approximation for the electric field leading to the model in Refs. [5,6] fails). Error bars (smaller than the dot size) are calculated as the maximum range of each peak energy distribution. Statistical errors in the linear fit to the model are obtained as the largest difference between parameters derived for the linear fits of the mean peak values, the mean values with maximal positive, and the mean values with maximal negative errors. Note the peaks below 3 V are features of the Ag an MgO electronic structure, and they are not taken into account for the analysis of the workfunction.

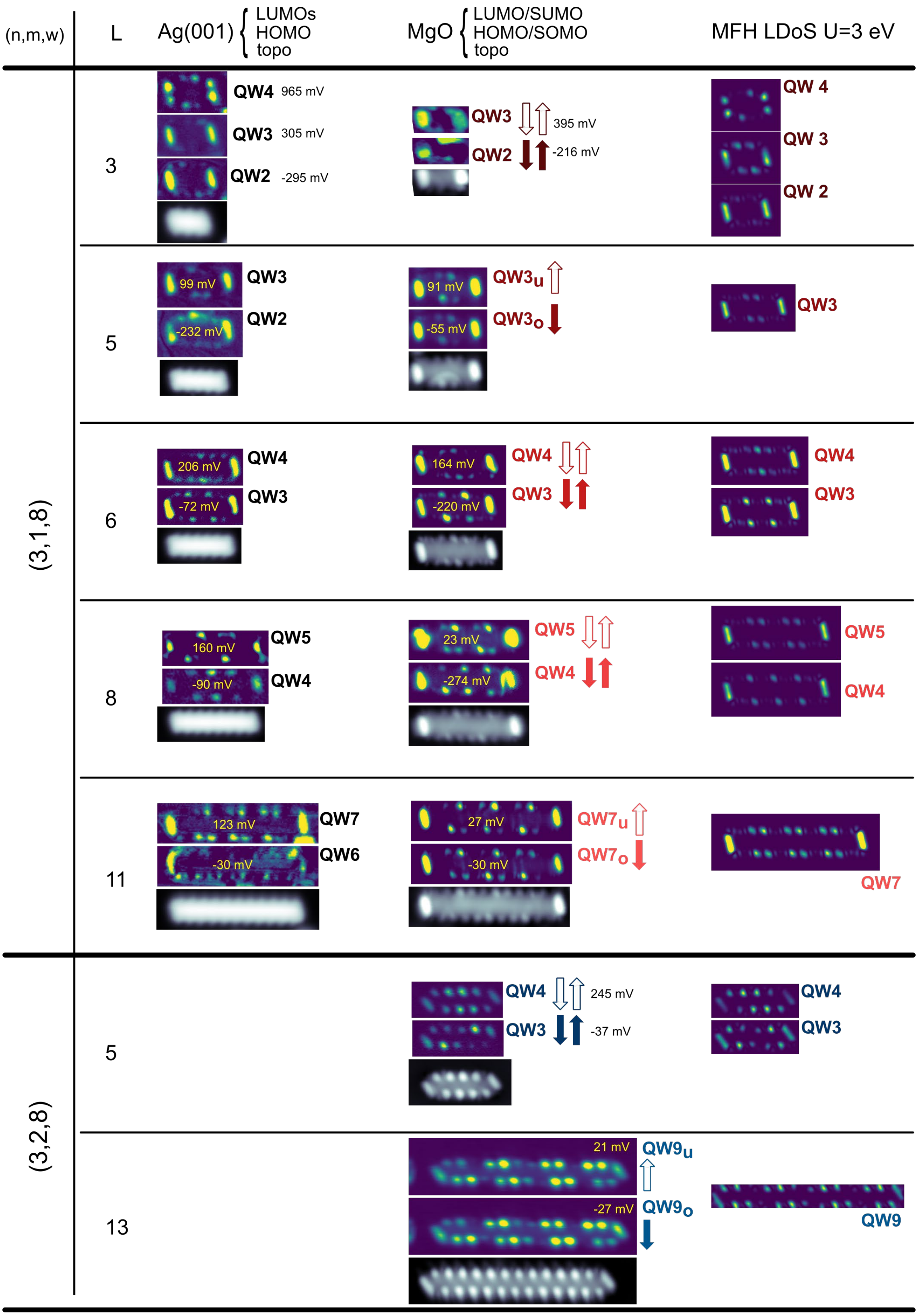

**Figure S8.-** Collection of experimentally mapped frontier states in Ag(001) and MgO. When the gap in MgO is much smaller than in Ag(001), the two frontier states have the same spatial distribution and thus the ground state becomes spin ½. The comparison with the calculated LDoS QW states with the same spatial distribution allows us to determine the GNRs charge state for each case. STM parameters for GNRs on MgO: constant height scans at the specified sample bias with lock-in modulation of 2 mV rms, except for $L$=3 for which we plot a constant current scan with regulation set point of 500 mV and 20 pA and modulation 4.4 mV rms. STM parameters for GNRs on Ag(001):

constant current maps at the specified sample bias for $L$=3, 5, 6, 11; constant current maps with regulation set point of -300 mV and 100 pA for $L$=8. All topographies (grey scale images) are taken at 500 mV sample bias. Labels and color code of schematic spin states are the same as in Figs. 3 and 4 of the main text.

| (n,m,w) | L / pu | Ag(001) | | | $MgO_{ML}$/Ag(001) | | | $\Delta E_{MgO-Ag}$ |
|---|---|---|---|---|---|---|---|---|
| | | Last occ. QW ($k^{th}$) | INT(#e) | e/PU (≃) | Last occ. QW ($k^{th}$) | #e | e/PU | Shift $(k+1)^{th}$ QW / eV |
| 318 | 3 | 2 | 6 | 2 | 2 | 6 | 2 | +0.075 |
| | 5 | 2 | 6 | 1.2 | **½ 3** | 7 | 1.4 | -0.1 |
| | 6 | 3 | 8 | 1.33 | 3 | 8 | 1.33 | -0.07 |
| | 8 | 4 | 10 | 1.25 | 4 | 10 | 1.25 | -0.137 |
| | 10 | 4 | 12 | 1.2 | -- | -- | -- | -- |
| | 11 | 6 | 14 | 1.27 | **½ 7** | 15 | 1.36 | -0.123 |
| | 12 | 5 | 14 | 1.16 | -- | -- | -- | |
| 328 | 5 | | | | 3 | 8 | 1.6 | |
| | 13 | -- | -- | -- | **½ 9** | 19 | 1.46 | |

**Table S1.-** List of approximate charge states on Ag(001), integer charge states on $MgO_{ML}$ last occupied QW states and shift of the first fully unoccupied QW state. The order of the QW state has been deduced by comparing experimental dI/dV maps of the frontier orbitals with the theoretical LDoS. The ½ symbol indicates that the referred QW state is singly occupied.

Figure 85 collects the experimental LDoS maps of the QW resonances for all the GNRs that were successfully characterized on the $MgO_{ML}$, together with the corresponding states as they appear on the bare Ag(001). A comparison with the simulated LDoS using the MFH model with $U$=3 eV (see Methods at the main article body), shown in the right hand column, allows us to extract the quantum number associated to the QW order, and thus, the charge state (as explained at the discussion of Fig. 3 of the main text). The so deduced charge states and resulting doping densities for GNRs on Ag and $MgO_{ML}$ are given in Table S1.

The first unoccupied QW state experiences a shift in energy of approximately -100 meV when moving the GNR ($L$>3 PU) from the Ag to the $MgO_{ML}$ (schematically represented by grey dashed lines in Fig. 3 of the main manuscript). Surprisingly, this shift and the concomitant charge transfer from the underlying metal, occurs in a step-wise manner as more PU units are positioned over the MgO during the insertion. This is shown in Fig. S9, where we provide *dI/dV* spectra of 4 different GNRs in different stages of the lateral atomic manipulation process.

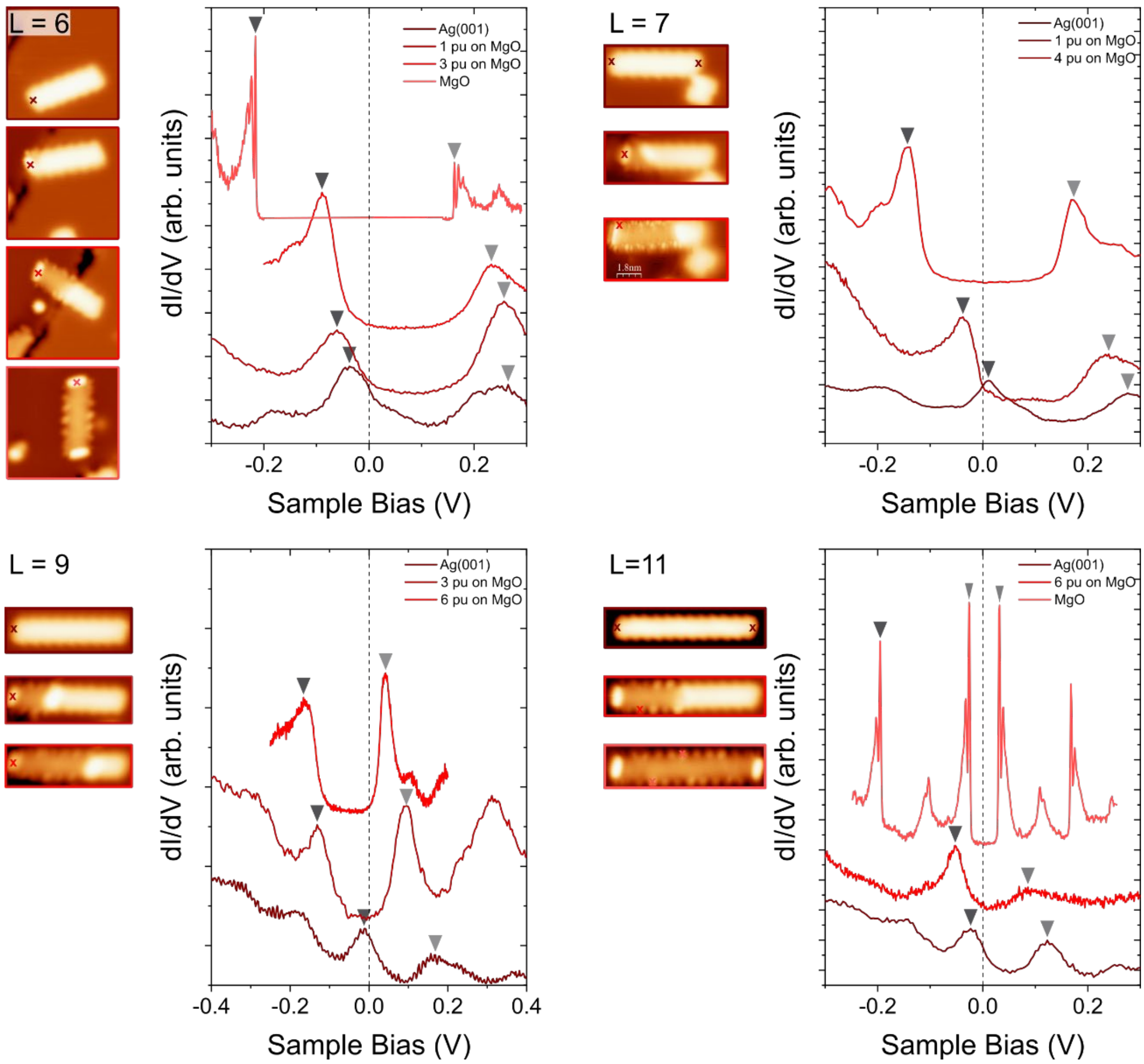

**Figure S9.-** Gradual insertion of (3,1,8)-GNRs on MgO and concomitant rigid downshift of the QW states. Horizontal image size: *L*=6 → 8 nm; *L*=7→ 9 nm; *L*=9→9 nm ; *L*=11→11.5 nm. The correlations induced splitting of the edge state only appears for fully decoupled ribbons. When some portion of the GNR remains in contact with the Ag(001), the n-doping takes place gradually depending on the length inserted onto $MgO_{ML}$ and metal, but the e-e interaction is partially quenched. This explains why previous attempts with (1,0,6)-GNR/Au(111) partially intercalated with NaCl [7] did not manifest the true discretization observed on our $MgO_{ML}$ patches.

## III. Identification of the SPT state and charge neutrality point of GNRs.

The charge neutral GNRs are characterized by a half-filled SPT state [8], and therefore the binding energy of this state will give the charge neutrality energy level for the charged case. The simulated DOS of a 11PU (3,1,8)-GNR with the observed charge state (15 e-) is shown in Extended Data Fig. 2). The QW resonances with low index (close to the SPT states) are very close in energy (they correspond to the first LUMOs of the neutral GNR, which originate from discretization of the parabolic conduction band of the edge states), and merge into a broad peak as a consequence of the overlap with vibrational resonances and the small intrinsic linewidth of all the spectroscopic features. This broad band is peaked at the value of the SPT end states, which lie at -410 meV in the simulated DOS and -650 meV in the experimental dI/dV spectra. The downshift of the SPT end states is due to the 500 meV change of the chemical potential, which is the value yielded by our calculation -based on equation (2)- to fill the the GNR with 15 electrons.

The experimental part of Extended Data Fig. 2 shows a dI/dV spectrum in a wide bias region, showing a set of multiple peak features. Through the analysis of the dI/dV maps and comparing with the corresponding ones in MFH simulations, we can safely identify the half occupation of QW7 and states around this orbital (QW6-QW8). We note that while these peaks may change their intensity on different GNR positions, they are always detected at about the same energy position. The peaks are very sharp and, therefore, are very sensitive to the variations of the electrostatic potential at MgO surface (e.g. see the parabolic shape of the line spectra in Fig. 2 of the manuscript). This causes small discrepancies in the intensity of the maps with respect to DOS maps, which can only be corrected applying the FD-STS method explained in Supplementary Section V and Fig. S12a-b. However, the number of nodal planes hints to the right identification of each corresponding QW state.

Below these energy (and, also, above), a sharp dI/dV peak (marked with asterisks in Exteded Data Fig. 2) changing abruptly its energy with the tip position is attributed to charging events mediated by point defects of the MgO substrate (see Fig. S12e-k). Most probably, that peak is also contributed by QW4 and the vibrational resonances of QW5, which can be also shifted differently in each charge state. Consequently, the dI/dV maps in this energy range of -400 to -600 meV show features that cannot be identified uniquely by a single QW state, although the dI/dV intensity pattern of at -500 meV is very close to that of the simulated QW3, and the characteristic intensity at the acute corner of the ribbon in SPT states is visible in the -662 mV map.

The charge neutrality point can be probed at even larger negative: we observe an increase of dI/dV attributed to the broad spectral band predicted by MFH simulations. According to the above discussion, this band, peaked around -0.65 eV, is the indication of the neutrality point of the GNR. This state appears ~200 meV lower than predicted by the model. This is probably due by the combined action of downwards shifts caused by i) the charging events, ii) a small potential drop that may occur across the MgO layer, and iii) some renormalization of levels in the charged molecule. Furthermore, our MFH model ubiquitously uses 3 eV as on-site Coulomb repulsion energy. Fine tuning of this parameter would allow us to adjust the chemical potential in equilibrium. Still, the simulated long-range spectrum is consistent with our identification of frontier orbitals, and our parametrization captures all essential Physics of the system.

Exteded Data Fig. 1 includes a set of similar maps and spectra for a 14PU (3,1,8)-GNR on Ag(100). In this case, QW8 and QW9 are clearly identified as Frontier states, consistent with an electron charge between 16-18 electrons. Other QW states (e.g. QW10 and QW3) can also be identified from the map's comparison as no charging events take place on the metal. Regarding identification of lower QW states and the neutrality point, we note that on the Ag(100) surface the QW peaks are wider. The lower QW states merge in a broad state peaked at 500 meV below zero, which is pictured in the MFH simulations as the position of the SPT state of the charge neutral case. This value is also in line with the predicted shift of our model for a charge transfer of 1.2 $e^-$/PU of about -400 meV.

# IV. Supplementary Theoretical Methods.

## IV.1. Charging model within the grand canonical ensemble formalism.

It is useful to describe the Ag(100)/$MgO_{ML}$/GNR system of our experiment as a plate capacitor model formed by components with different electron affinity ($E_a$). This model is validated in Ref. [9]. The MgO layer acts as electronic decoupler, that hinders the wavefunction overlap between metal and molecular states, and stabilizes integer states in the GNR (rather than fractional, as would happen for direct GNR adsorption on the metal). At the same time, the MgO layer reduces the substrate's work function [10–12]. To reach chemical equilibrium with the metal, the GNR becomes charged by an integer amount of electrons that tunnel from the metal through the MgO layer in response to their difference in electron affinities. The charging of the GNR simultaneously causes an electric field across the MgO that opposes the charging and partially compensates the difference in electron affinities. This component has been described in previous works as an interface dipole that causes deviation from pure vacuum-level alignment [9]. Therefore, in equilibrium these processes are reflected by a shift of the GNR's chemical potential $\mu$ with respect to the neutral case level $\mu_0$ (i.e. by $\Delta\mu = \mu - \mu_0$), which is smaller than the difference in electron affinities owing to the energy $U_d$ stored in the MgO capacitor by the built-up dipole. This is reflected in the following expression

$$\Delta\mu = \mu - \mu_0 = E_a - \Phi - U_d \quad (1)$$

The value of $\Delta\mu$ determines the charge state of a GNR of length $L$ in equilibrium. The two types of chiral GNRs used in the experiment have a non-trivial topological band structure and, therefore, two symmetry-protected topological (SPT) end states [8]. For the neutral GNR, the two SPT states are half occupied, and therefore the neutral level $\mu_0$ is referenced by their binding energy (Fig. 5a main article). Upon electron charging, the chemical potential $\mu$ will lie either between two LUMO levels of the molecule for even charge (i.e. the closed-shell configuration, singlet spin state), or between SOMO and SUMO levels, for the case of odd charge state (open shell, doublet spin state).

To simulate how the GNR charges as a function of length, we describe the system in the grand canonical ensemble and calculate the mean number of excess electrons $\langle q \rangle$ for any given chemical potential $\mu$ using the relation

$$\langle q(\mu, L, T) \rangle = -k_B T \frac{dZ(\mu,L,T)}{d\mu} \qquad (2)$$

This expression provides a charge value as a function of μ which can be compared with experimental results for every GNR. For a GNR with a given length $L$ and chirality, at the temperature of our experiment (between 1.2 and 4.3 K), we can calculate the grand canonical partition function $\mathcal{Z}(\mu, L, T)$ following the expression

$$\mathcal{Z}(\mu, L, T) = \sum_j exp\left(\frac{N_jT - E_q}{k_BT}\right) = \sum_q exp\left(\frac{j\mu - E_0 - E_q}{k_BT}\right) \quad (3)$$

where $N_j$ is the total number of indistinguishable electrons, and $E_j$ the internal energy of each respective charge state $j$ obtained from MFH simulations (see Methods main text). No overcounting correction is included as we consider each microstate $j$ only once. We note that at cryogenic temperatures, the entropic contribution to the grand potential can be neglected and $\langle q \rangle$ can be obtained equivalently by simply minimizing the right hand side of (3), instead of the total energy. We limit our model to the spin-restricted case, as the mean spin densities on the edge are energetically unstable against the emergence of nodal planes, and thus determining the correct ground state becomes ambiguous for large (3,*n*,8)-GNRs doped into the chiral edge band.

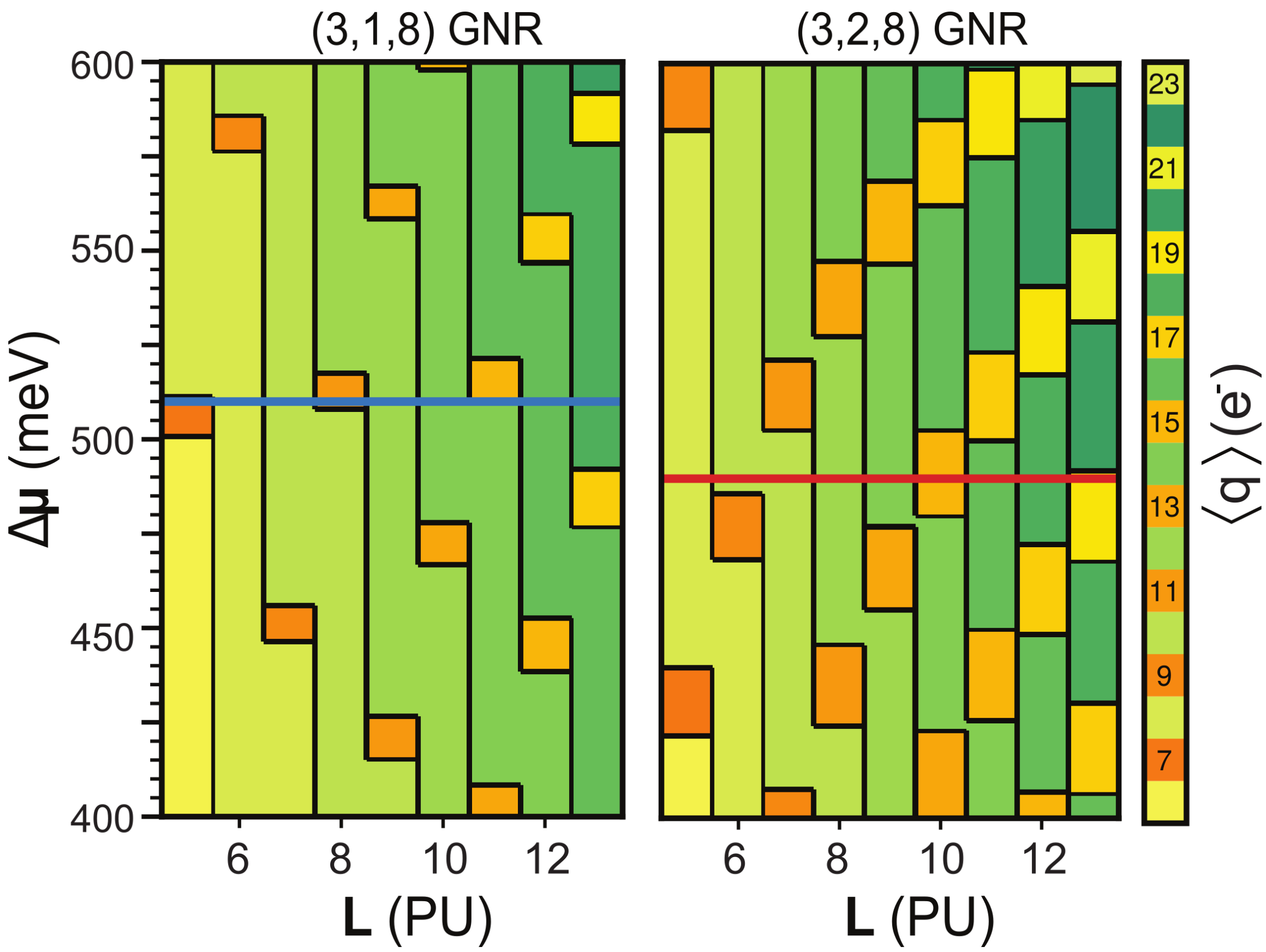

**Figure S10:** Contour plots of the calculated excess charge for (3,1,8)- (left) and (3,2,8)-chGNRs (right). Even and odd charge states are represented by different colour scales. Horizontal lines represent the best-fit values for $\Delta\mu$ to match the experimental charging pattern (see Fig. 5b of the main text).

Figure S10 maps the predicted charge patterns as a function of GNR's length as the chemical potential shifts $\Delta\mu$ from 400 to 600 meV. The map reflects the increase in charge with length for both chiralities, and the peculiar odd-even (orange-green) charge pattern that depends on the value of $\Delta\mu$. Comparing the charging patterns with the experimental charge states of the GNRs (Fig. 4 of the main text), an estimated value for $\Delta\mu$, the chemical potential shift, is determined and represented in Fig. S10 by horizontal lines. The theoretical pattern and its comparison with experiments is plotted in Fig. 5b of the main text. The only deviation occurs for the ribbon (3,1,8)-GNR with $L$=8, which appears with even electron occupation ($q$=10) in the experiment, while in the map is expected to have 11 elementary charges. As discussed in Section V, small variation of $\Delta\mu$ caused by local defects in the MgO layer can explain this discrepancy.

### IV.2. Modeling vibrational modes of chiral GNRs on MgO.

We performed calculations of the vibrational modes of chiral GNRs using atomistic model potentials with the GULP package (see Gale and Rohl, Molecular Simulation 29, 291 (2003)) [13]. In detail, we modelled the GNR itself by an accurate potential for hydrocarbons [14] while the molecule-surface interaction is described by simple Lennard–Jones 12-6 potentials between pairs of atoms with a varying strength $E_{LJ}$. We treat the MgO film as a rigid surface.

The total density of vibrational modes (each one represented with smearing factor of 3 meV) is shown in the negative scale of Supplementary Figure S11 for different number of precursor units (L) and interaction strength with the model substrate ($E_{LJ}$). To nail down the origin of these modes, we computed the out-of-plane density of vibrational states as the norm of the z-components of their displacement eigenstates, which is shown in the positive scale of the figure.

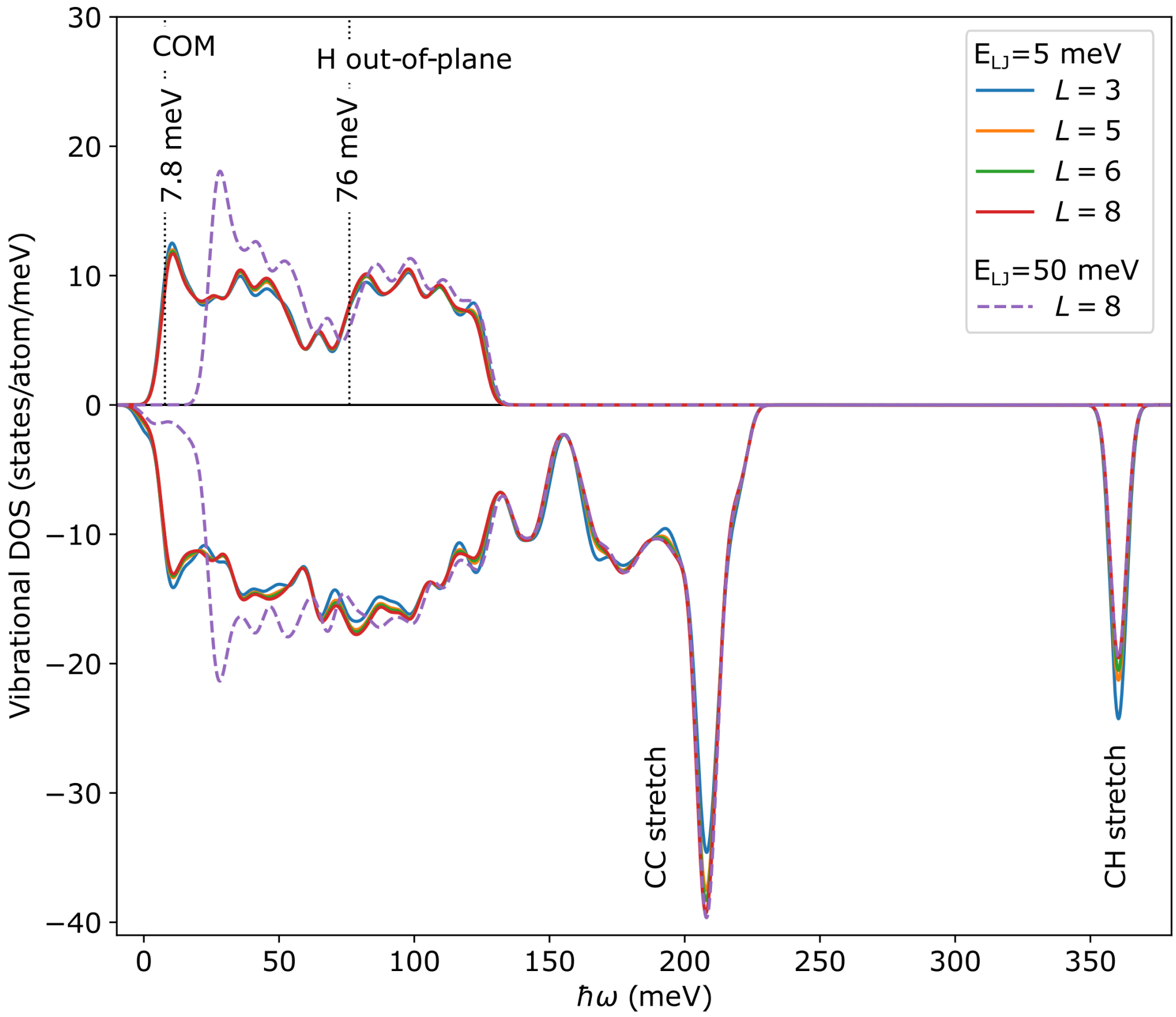

**Figure S11.-** Vibrational density of states for (3, 1, 8)-GNRs adsorbed on a model MgO film for which the atomic interaction is described by a 12-6 Lennard-Johns potential with strength given by the energy scale of $E_{LJ}$. The negative DOS corresponds to the total vibrational states. The positive DOS represent out-of-plane vibrations. The discrete spectrum is broadened by a Gaussian function with a smearing of 3 meV. Dotted lines mark the energy of the experimentally obtained most intense vibrational peaks (see Fig. S6), which are ascribed to vertical vibrations of the GNR center of mass (COM, thereby the ribbon as a whole) and to the concerted out-of-plane vibration of the hydrogen atoms.

The first peak in the spectrum at lowest energy corresponds to an out-of plane translation of the molecule's center of mass (COM), thereby an external vibration. Its energy scales with $E_{LJ}$, and we therefore use this parameter to match the observed mode energy at 7.5 meV (see experimental vibrational resonances in Fig. S6b-c). This indicates that the concerted movement of the whole GNR with respect the MgO surface is the most probable mode describing the 7.5 meV FC resonance. In the out-of-plane part of the vibrational spectrum, a secondary band emerges in the energy range 70-130 meV. These vibrational modes include those with predominantly H-atoms moving out-of-plane. The onset of this secondary band is very similar to the 76 meV peak observed in the experiments (see Fig. S6). As shown in Fig. S11, these features are rather insensitive to the length of the GNR, in agreement with the experimental data.

At higher energies there are additional vibrational modes in the region corresponding to the C-C stretching modes, which disappear when only vertical motion is considered (because they do not have out of plane components). As shown in Fig. S11, these internal modes do not react to the GNR-MgO interaction potential strength, whereas the external modes are clearly affected, outlining their different nature.

In summary, this simple analysis suggests that the two satellite peaks experimentally observed in dI/dV may be ascribed to the excitation of out-of-plane vibrational modes, and more specifically to the external out-of-plane COM vibration and to the out-of-plane H-atom motion, respectively.

## V. Electrostatic gating of chiral graphene nanoribbons on MgO.

As shown in Fig. S12a-c and Fig. 2 of the main article, when the tip moves laterally along the GNRs on MgO, we often observe a rigid shift of all energy levels, $\Delta E_p$ (which would not be visible in the case of molecular orbitals spanning in an energy range wider than the maximum $\Delta E_p$~20 meV). In several cases we have been able to find a direct relationship between the $\Delta E_p$ and the lateral tip distance to point defects of the MgO that can be detected either below the GNR or next to it. In Fig. S12 we present an example of this effect for the (3,1,8)-GNR with $L$=11 discussed in the main text.

If we apply the Feature Detection STS method [15] to compose a map of the area where a spectroscopic feature with the shape of the QW resonances (i.e., a peak with FWHM of 1.5 mV, see Fig. S5) exists, the intensity pattern of the QW edges is evenly distributed at the locations predicted by MFH simulations (Fig. S12d). This is in contrast with the case of constant height $dI/dV$ maps at a given energy of the same GNRs, where some lobes of the QW state appear brighter than others (see for example Fig. S8 or Fig. 3b). The reason is that as the tip position $\vec{\boldsymbol{r}} = (x, y, z)$ departs from the defect at $\vec{\boldsymbol{r}}_0 = (x_0, y_0, z_0)$, the electrostatic potential energy at the defect $U_e(\vec{\boldsymbol{r}} - \vec{\boldsymbol{r}}_0)$ varies. Indeed, it has been previously shown that the electric field in the tip-sample gap can modify controllably the electric polarization of thin insulating layers [16], specially next to point defects and vacancies. This gives rise to a variation of the local value of $\Delta\mu(\vec{\boldsymbol{r}} - \vec{\boldsymbol{r}}_0)$ which, in the parallel plane approximation, will be proportional to $U_e$. This effect can be viewed as an effective gating of the molecular states of the GNRs.

The location of the defect can be pulled out from measurements of the tip induced charging resonances of the defect states [4,17]. They manifest as sharp ellipsoidal rings in the $(x, y)$ plane enclosing the region for which $U_e(\vec{\boldsymbol{r}} - \vec{\boldsymbol{r}}_0)$ is large enough as to charge (positively or negatively) the defect at a constant $z - z_0$. Fixing $V_b$ at a large value of 0.5 V, the size of the charging ring must depend linearly on the tip sample distance, or equivalently, logarithmically on the current set point $I_t$ before opening the feedback (see Figs. S12e-k). As shown in Fig. S12j, we can determine accurately the defect position as the centre of an ellipse fitting the charging ring for different tip heights.

Alternatively, the defects can be found in atomically resolved images of the MgO patch after removing the GNR under study, as illustrated in Fig. S13. Here, we show how the QW states of the (3,1,8)-cGNR with $L$=5 PU can be shifted in energy by changing the distance to a point defect on $MgO_{ML}$. The GNR in position γ lies over a defect free region of the MgO, and displays a correlation gap between singly occupied frontier states –as predicted by the model discussed in supplementary theoretical methods-. In contrast, when the GNR lies over a point defect (position α) or next to the MgO bilayer edge (position β), the 3rd QW state (see Fig. S8) is shifted in energy above the Fermi level and becomes fully unoccupied, yielding another charge ($q$=6 electrons instead of 7) and total spin ($S$=0 instead of $S$=1/2).

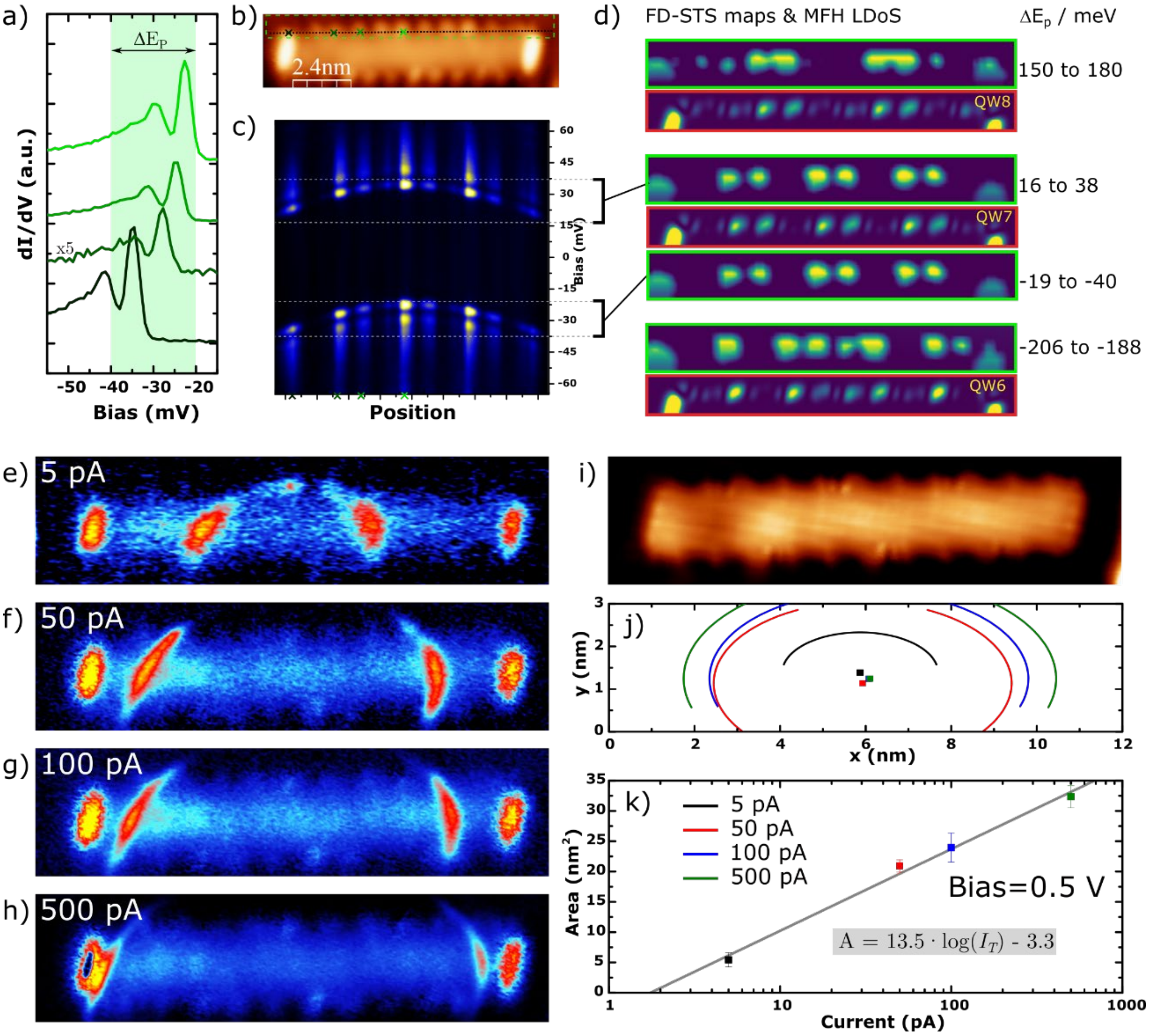

**Figure S12.- Gating energy levels by lateral tip positioning. a-c)**: Shift in energy along the chiral edge of the (3,1,8)-GNR with $L$=11. (STS set point: 0.5 V, 200 pA, $V_{mod}$=1 mV). a) STS spectra recorded at different positions of the chiral edge. b) STM topography image of the $L$ = 11 (3,1,8)-GNR including crosses where STS spectra shown in (a) are recorded. Image set point: 0.5 V, 100 pA. c) Stack plot of $dI/dV$ versus position recorded on the chiral edge of the GNR shown in b (dotted line). **d)** Feature-Detection STS [15] intensity maps tuned to a peak with linewidth ≤2 mV in the energy range $\Delta E_p$ indicated on the right column. Due to the high spatial and energy resolution required for this analysis, we only studied the frame enclosed by the dashed green rectangle in (b). Each experimental map is compared with the corresponding theoretical LDoS of the QW states obtained from the MFH model with $U$=3 eV and $q$=15 electrons (see theoretical methods at the main text) for the $L$ = 11 (3,1,8)-GNR. **e-k)** Determination of the location of point defects in $MgO_{ML}$ by high bias mapping. e-h) Series of constant height $dI/dV$ maps recorded at 0.5 V with tunnelling current set points from 5 to 500 pA at the time of feedback opening over the ribbon centre ($V_{mod}$ = 4 mV). i) In-gap constant height STM image of the GNR shown in e-h ($V_b$=-2.5 mV). j) Fits (lines) and centers (squares) of the ellipses extracted from the $dI/dV$ maps. k) Area of the ellipses as a function of set point tunnelling current.

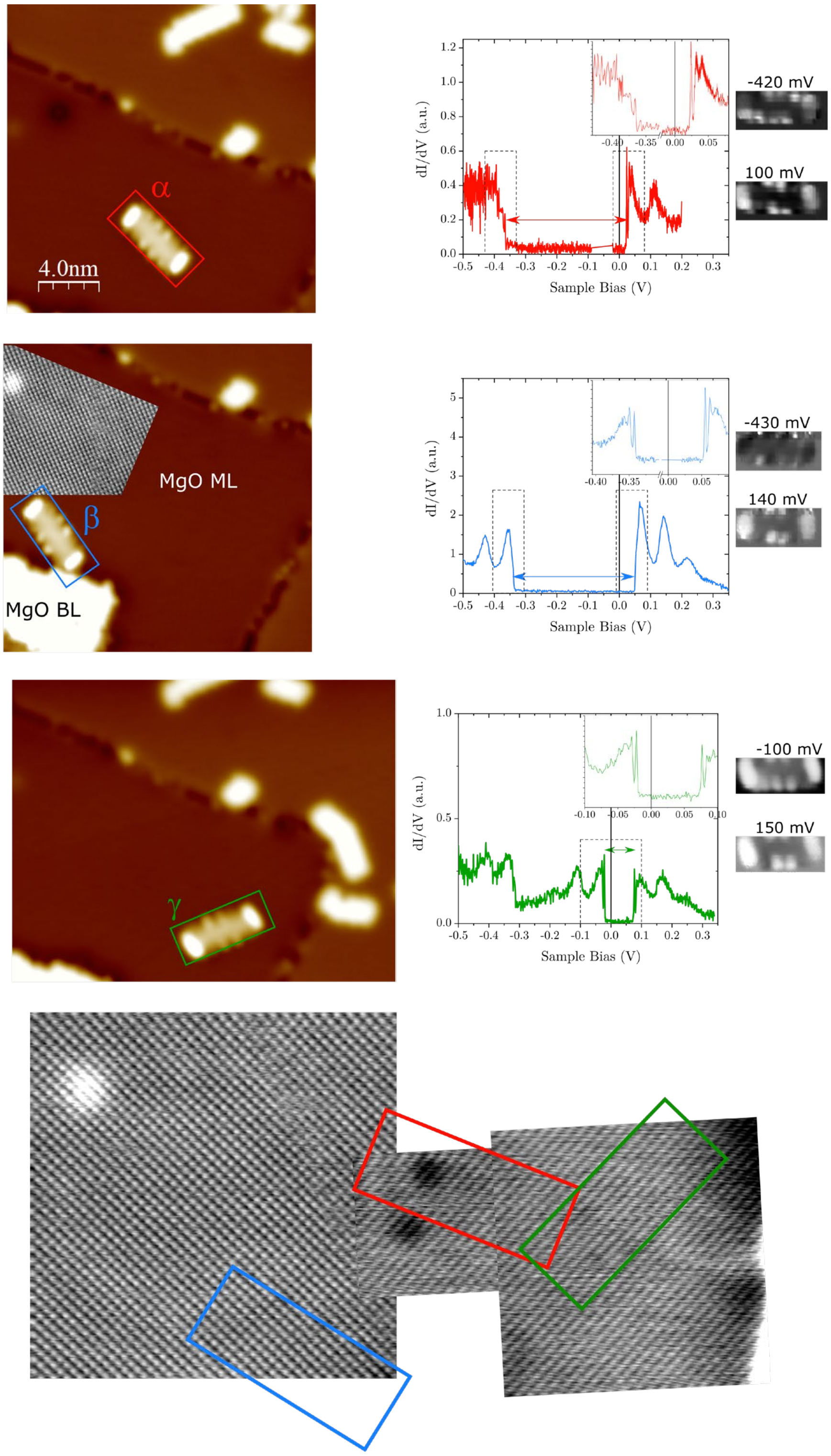

**Figure S13.- Gating electron occupancy by GNR position.** Data corresponds to the $L$=5 (3,1,8)-GNR. $V_{mod}$=1 mV and 0.5 mV r.m.s. for spectra in main panels and in the insets, respectively (except for the $\beta$ position, for which the main panel spectra is taken with $V_{mod}$=5 mV). The maps of the QW states are tunnelling current images taken at the referred energies but regulating at each pixel at 0.5 V. The three upper panels show the same GNR at three different positions over the MgO patch. As shown in the atomically resolved image of the substrate (grey color scale), in position α the GNR is over a dark spot and next to the bright defect on MgO. It features a normal quantization gap and $q$=6 electrons and therefore total spin $S$=0. Position β is next to the MgO bilayer (BL) island, and the GNR has same charge/spin state as in α. However, at the γ position, farther away from all types of defects, the GNR

is over a defect free region of the MgO, and now exhibits a much smaller gap, as corresponds to the correlations gap. In this latter case, the current map integrating the resonance of the frontier states are the same at both sides of the Fermi level, and thus it has $q$=7 electrons (odd occupancy) and total spin $S$=1/2.

# VI. Supplementary References

[1] A. Berdonces-Layunta, F. Schulz, F. Aguilar-Galindo, J. Lawrence, M. S. G. Mohammed, M. Muntwiler, J. Lobo-Checa, P. Liljeroth, and D. G. de Oteyza, Order from a Mess: The Growth of 5-Armchair Graphene Nanoribbons, ACS Nano **15**, 16552 (2021).

[2] G. Reecht, N. Krane, C. Lotze, L. Zhang, A. L. Briseno, and K. J. Franke, Vibrational Excitation Mechanism in Tunneling Spectroscopy beyond the Franck-Condon Model, Phys. Rev. Lett. **124**, 116804 (2020).

[3] J. van der Lit, M. P. Boneschanscher, D. Vanmaekelbergh, M. Ijäs, A. Uppstu, M. Ervasti, A. Harju, P. Liljeroth, and I. Swart, Suppression of electron–vibron coupling in graphene nanoribbons contacted via a single atom, Nat Commun **4**, 2023 (2013).

[4] M. S. G. Mohammed, L. Colazzo, R. Robles, R. Dorel, A. M. Echavarren, N. Lorente, and D. G. de Oteyza, Electronic decoupling of polyacenes from the underlying metal substrate by sp3 carbon atoms, Commun Phys **3**, 159 (2020).

[5] J. Bono and R. H. Good, Conductance Oscillations in Scanning Tunneling Microscopy as a Prob of the Surface Potential, Surface Science **188**, 153 (1987).

[6] O. Yu. Kolesnychenko, Yu. A. Kolesnichenko, O. I. Shklyarevskii, and H. van Kempen, Field-emission resonance measurements with mechanically controlled break junctions, Physica B: Condensed Matter **291**, 246 (2000).

[7] S. Wang, L. Talirz, C. A. Pignedoli, X. Feng, K. Müllen, R. Fasel, and P. Ruffieux, Giant edge state splitting at atomically precise graphene zigzag edges, Nature Communications **7**, 11507 (2016).

[8] J. Li, S. Sanz, N. Merino-Díez, M. Vilas-Varela, A. Garcia-Lekue, M. Corso, D. G. de Oteyza, T. Frederiksen, D. Peña, and J. I. Pascual, Topological phase transition in chiral graphene nanoribbons: from edge bands to end states, Nat Commun **12**, 5538 (2021).

[9] P. Hurdax, M. Hollerer, P. Puschnig, D. Lüftner, L. Egger, M. G. Ramsey, and M. Sterrer, Controlling the Charge Transfer across Thin Dielectric Interlayers, Adv. Mater. Inter. **7**, 2000592 (2020).

[10] M. Hollerer, D. Lüftner, P. Hurdax, T. Ules, S. Soubatch, F. S. Tautz, G. Koller, P. Puschnig, M. Sterrer, and M. G. Ramsey, Charge Transfer and Orbital Level Alignment at Inorganic/Organic Interfaces: The Role of Dielectric Interlayers, ACS Nano **11**, 6252 (2017).

[11] G. Witte, S. Lukas, P. S. Bagus, and C. Wöll, Vacuum level alignment at organic/metal junctions: "Cushion" effect and the interface dipole, Appl. Phys. Lett. **87**, 263502 (2005).

[12] H.-J. Freund and G. Pacchioni, Oxide ultra-thin films on metals: new materials for the design of supported metal catalysts, Chem. Soc. Rev. **37**, 2224 (2008).

[13] J. D. Gale and A. L. Rohl, The General Utility Lattice Program (GULP), Molecular Simulation **29**, 291 (2003).

[14] D. W. Brenner, O. A. Shenderova, J. A. Harrison, S. J. Stuart, B. Ni, and S. B. Sinnott, A second-generation reactive empirical bond order (REBO) potential energy expression for hydrocarbons, J. Phys.: Condens. Matter **14**, 783 (2002).

[15] J. Martinez-Castro et al., Disentangling the electronic structure of an adsorbed graphene nanoring by scanning tunneling microscopy, Commun Mater **3**, 57 (2022).

[16] J. Martinez-Castro, M. Piantek, S. Schubert, M. Persson, D. Serrate, and C. F. Hirjibehedin, Electric polarization switching in an atomically thin binary rock salt structure, Nature Nanotechnology **13**, 19 (2018).
[17] C.-L. Song, Y.-P. Jiang, Y.-L. Wang, Z. Li, L. Wang, K. He, X. Chen, X.-C. Ma, and Q.-K. Xue, Gating the charge state of single Fe dopants in the topological insulator Bi_{2}Se_{3} with a scanning tunneling microscope, Physical Review B **86**, (2012).